\documentclass[aps,prx,twocolumn,showpacs,superscriptaddress,longbibliography]{revtex4-1}

\usepackage[colorlinks=true,linkcolor=blue,citecolor=red]{hyperref}
\usepackage{graphicx}
\usepackage{dcolumn}
\usepackage{bm}
\usepackage{amssymb} 
\usepackage{mhchem}
\usepackage{amsmath}
\usepackage{xcolor} 
\usepackage{makecell} 
\usepackage{changes}
\usepackage{cancel}

\def\bk{\boldsymbol{k}}
\def\cc{c^{\dagger}}
\def\ca{c^{\phantom{\dagger}}}

\begin{document}

\title[Thermal dynamics and electronic temperature waves in layered correlated materials]{
Thermal dynamics and electronic temperature waves in layered correlated materials} 
\author{Giacomo Mazza}
\email{giacomo.mazza@unige.ch}
\affiliation{Department of Quantum Matter Physics, University of Geneva, Quai Ernest-Ansermet 24, 1211 Geneva, Switzerland}
\author{Marco Gandolfi}
\affiliation{CNR-INO, Via Branze 45, 25123 Brescia, Italy}
\affiliation{Department of Information Engineering, University of Brescia, Via Branze 38, 25123 Brescia, Italy}
\author{Massimo Capone}
\affiliation{CNR-IOM Democritos National Simulation Center and Scuola Internazionale Superiore di Studi Avanzati (SISSA), Via Bonomea 265, 34136 Trieste, Italy}
\author{Francesco Banfi}
\email{francesco.banfi@univ-lyon1.fr}
\affiliation{FemtoNanoOptics group, Universit\'e de Lyon, CNRS, Universit\'e Claude Bernard Lyon 1,Institut Lumie Mati\`ere, F-69622 Villeurbanne, France}

\author{Claudio Giannetti}
\email{claudio.giannetti@unicatt.it}
\affiliation{Dipartimento di Matematica e Fisica, Universit\`a Cattolica del Sacro Cuore, Via Musei 41, I-25121 Brescia, Italy}
\affiliation{Interdisciplinary Laboratories for Advanced Materials Physics (I-LAMP), Universit\`a Cattolica del Sacro Cuore, Via Musei 41, I-25121 Brescia, Italy}

%

\begin{abstract}
We explore layered strongly correlated materials as a platform to identify and control 
unconventional heat transfer phenomena.
We demonstrate that these systems can be tailored to sustain a wide spectrum of heat transport regimes, ranging from ballistic, to hydrodynamic all the way to diffusive. Within the hydrodynamic regime, wave-like temperature oscillations are predicted up to room temperature. All the above phenomena have a purely electronic origin, stemming from the existence of two components in the electronic system, each one thermalized at different temperatures. The interaction strength can be exploited as a knob to control the different thermal transport regimes.
The present results pave the way to transition-metal oxide heterostructures as building blocks for nanodevices exploiting the wave-like nature of heat transfer on the picosecond time scale.
\end{abstract}

\maketitle
\section{Introduction}
Understanding the mechanism of heat transfer in nanoscale devices remains one of the greatest intellectual challenges in the field of thermal dynamics, by far the most relevant under an applicative standpoint \cite{Chen2005,volz2016,li2012,luo2013,cahill2014nanoscale}. When thermal dynamics is confined to the nanoscale, the characteristic timescales become ultrafast, engendering the failure of the general assumptions on which the conventional description of energy propagation relies. 

The capability to access ultrafast thermal dynamics recently gave access to striking phenomena that take place in materials at the nanoscale before complete $\textit{local}$ energy equilibration among heat carriers is achieved. For instance, non-Fourier heat transport regimes have been reported for hot spots dimensions inferior to the phonon mean free-path \cite{Siemens2010,Minnich2011,Johnson2013}, in which energy is ballistically carried point to point, or have been engineered via nano-patterning of dielectric substrates \cite{Hoogeboom2015,chen2018,Frazer2019}.
As a consequence of the existence of two non-thermal populations, wave-like thermal transport, often referred to as second sound \cite{Guyer1966,Beck1974}, has been predicted in graphene, both in the frame of microscopic \cite{Lee2015,ding2018,cepellotti2015,Li2019_PRB} and macroscopic models \cite{Gandolfi2019}. Temperature wave-like phenomena have been recently observed at high temperatures in graphene \cite{Huberman2019} and 2D materials \cite{Zhang2020} on sub-nanosecond timescales and scheme for their coherent control have been proposed \cite{Gandolfi2020}. 
So far most of the effort has been devoted to \textit{phononic} non-Fourier heat transport \cite{Cepellotti2017, Torres2019,Li2019_PRB,Huberman2019,Machida2020}, where, only recently, a theoretical framework, covering on equal footing Fourier diffusion, hydrodynamic propagation, and all regimes in between, has been proposed \cite{simoncelli2020}. On the contrary, despite its applicative relevance, \textit{electronic} non-Fourier heat transport remains relatively unexplored \cite{Gandolfi2019,Gandolfi2017,Zhang2020}.

Quantum correlated materials offer a new platform to control \textit{electronic} nanoscale heat transfer. The strong electronic interactions give rise to emerging many-body properties, such as collective and decoupled diffusion of energy and charge \cite{Hartnoll2015,Lee2017}. Tuning the interaction strength thus opens the possibility to investigate novel electronic regimes with no counterpart in conventional weakly-interacting materials \cite{Tokura2017,Basov_NatMat_2017}.  

In this work, we propose layered correlated materials (LCM) as the ideal platform to access the entire 
spectrum of unconventional \textit{electronic} heat transport regimes.
We present a microscopic description of the non-equilibrium dynamics and electronic heat transfer phenomena occurring in LCM on ultrashort space- and time-scales triggered by an impulsive excitation. 
We show that on sub-picosecond timescales the electronic heat transfer is initially characterized by 
ballistic wave-front propagation, followed by an hydrodynamic regime, which eventually evolves into
conventional Fourier heat transfer on longer timescales. 
In the hydrodynamic regime, we predict that LCM may sustain temperature wave oscillations at 
THz frequencies and up to ambient temperature.

The present work rationalizes the microscopic interactions underlying unconventional electronic heat transfer phenomena in LCM.
Our findings enlarge the functionalities of quantum materials \cite{Tokura2017,Basov_NatMat_2017} to the realm of nanoscale heat transport \cite{Miranda_PRB2018}, beyond the case of radiative energy transfer \cite{cesarini2019,Ben-Abdallah_2014,Miranda_PRL2019}.
Under an applicative stand-point these results pave the way to novel paradigms in thermal device concepts and to artificial nano-engineered materials \cite{Gandolfi2020}. 

\section{The platform: Layered correlated materials}
\begin{figure}[t]
\includegraphics[width=\columnwidth]{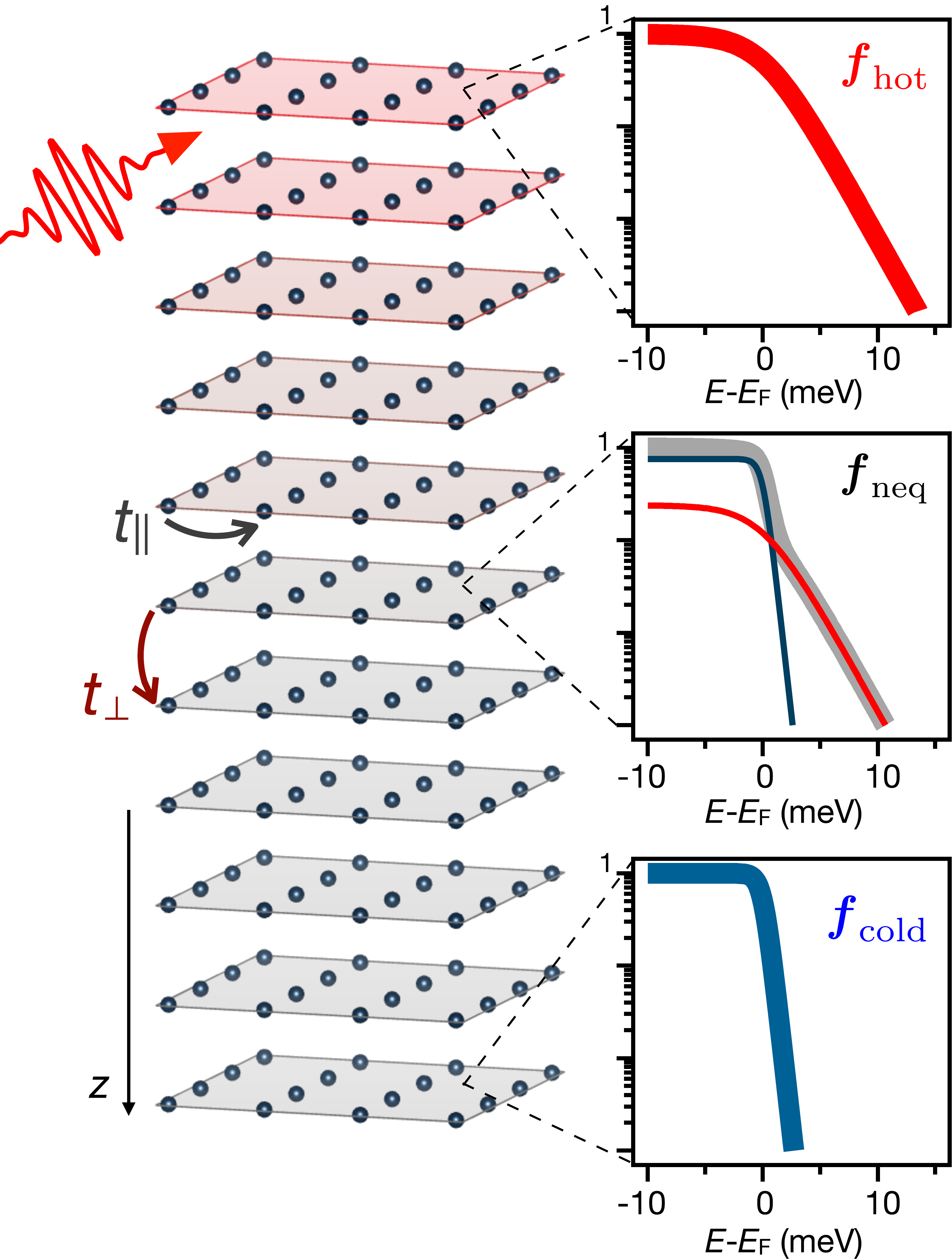}
\caption{\textbf{Setup}. Cartoon of the layered correlated material impulsively excited on the top surface by ultrafast light pulses. 
We assume that the excitation drives a fast thermalization of the electronic population establishing an electronic temperature 
$T_\mathrm{hot}$ on the topmost layers of the sample.
The right panels display the calculated transient electronic distributions at different depths. 
$f_\mathrm{hot}$ and $f_{\mathrm{cold}}$ are Fermi distribution
functions at temperatures $T_\mathrm{h}$ and $T$ respectively.
In the right-middle panel the non-equilibrium distribution is plotted in grey. 
In red and blue we report the {\it "hot"} and {\it "cold"} 
contributions to the non-equilibrium distribution function (see text). 
The transient temperature 
of the so-called "cold" electrons 
can be different from the initial temperature $T_{c0}$ (see text).}
\label{Figure_0}
\end{figure}

We consider an impulsive excitation on the surface of a LCM 
characterized by a strong local Coulomb interaction $U$ (see Fig. \ref{Figure_0}).
The interaction $U$ can drive fast local thermalization processes leading to the rapid 
build up of a hot intra-layer electronic temperature before relaxation via slower scattering paths takes place. 
At the same time, the interaction 
leads to heavier quasiparticles with enhanced effective mass $m^*$ and a reduced kinetic energy. 
As a consequence, energy propagation across the layers is expected to slow down for increasing $U$. Overall the interaction $U$
may thus act as a tuning parameter to control  the relative inter- and intra-layer energy 
exchange processes  in LCM.
Eventually, as the interaction increases, the two processes can effectively 
decouple, thus opening 
to novel electronic heat  transport regimes occurring on the ultra short space and time scales.

We investigate the possibility for unconventional heat transport regimes by focusing 
on the impulsive thermal dynamics of the layered single-band Hubbard model, which represents a 	
general framework for understanding  the effects of electronic interactions in a large family 
of correlated materials. 
The thermal dynamics is triggered by a sudden increase of the electronic temperature localized within the first few surface layers of  the LCM as can be achieved, for instance, by excitation with a femtosecond light pulse~\cite{Gandolfi2017}.
By tuning the interaction strength $U$ and the anisotropy of the system through the interlayer coupling $t_\perp$, 
we demonstrate that it is possible to control the energy transfer dynamics and explore three different heat transfer regimes: ballistic, hydrodynamic and Fourier-like. 

In order to contextualise the present concepts within the frame of real systems and to connect with the realm of technologically relevant materials, we focus on the correlated metal SrVO$_3$ (SVO). SVO is a paradigmatic representative of the wider class of correlated transition metal oxides (TMOs) and it has been proposed as a platform for  a wealth of potential technological applications ranging from ideal electrode materials \cite{Moyer2013}, to Mott transistors \cite{Zhong2015} and transparent conductors \cite{Zhang2015}. We argue that the degree of correlation of SVO, as measured by the interaction strength, is such that ballistic transport first, and wave-like thermal transport afterwords, are accessible on the sub-picosecond timescale. Our results, together with the possibility of heterostructuring TMO to atomic layer accuracy, promote these materials to ideal building blocks for nanothermal device architectures based on non-Fourier heat transport.

\subsection{The model}
In order to identify the intrinsic role of electronic correlations, we model the LCM by a simple single-band layered Hubbard model.
We do not include the interaction with phonons, which is however effective on longer timescales 
than the ones here addressed and does not significantly affect the present findings.
The Hamiltonian reads:
\begin{equation}
H = \sum_{n=1}^L h_n +\sum_{n=1}^{L-1} \tau_{n,n+1}
\label{eq:H_hubbard}
\end{equation} 
with 
\begin{equation}
h_n = \sum_{<i,j> \sigma} t_\parallel \cc_{i n \sigma} \ca_{j n \sigma} + U \sum_{i} n_{i n \uparrow} n_{i n \downarrow}  
\label{eq:h_layer}
\end{equation}
 and
 \begin{equation}
 \tau_{n,n+1} = \sum_\sigma t_\perp \cc_{i n \sigma} \ca_{i n+1 \sigma} + h.c.
 \label{eq:h_perp}
 \end{equation}
where $\cc_{i n \sigma}$ is a fermionic creation operator for an electron with spin $\sigma$
at the site $i$ belonging to the layer indexed by $n$, which ranges from 0 to $L$. 
$t_\parallel$ and $t_\perp$ represent, respectively, the intra- and inter-plane hopping amplitudes.
The sum in the in-plane hopping term  runs over pairs of nearest neighbouring  sites  and we introduce the number operator $n_{i n \sigma} = \cc_{i n \sigma} \ca_{i n \sigma}$.
We assume in-plane translational invariance so that we can introduce an in-plane momentum $\bk = (k_x,k_y)$ and recast 
$h_n = \sum_{\bk \sigma} \epsilon({\bk}) \cc_{\bk n \sigma} \ca_{\bk n \sigma} + U \sum_{i} n_{i n \uparrow} n_{i n \downarrow} $
with $\epsilon({\bk}) = -2 t_\parallel (\cos(k_x a) + \cos(k_y a))$ and $a$ the lattice spacing. 
 We fix the chemical potential in order to have an average occupation of one electron
per site (half-filling) corresponding to the perfect particle-hole symmetric case.
As a consequence, the total number of electrons per layer is conserved during the dynamics. 
This choice allows us to model energy transport in the absence of mass and charge transport.

We study the non-equilibrium dynamics in the frame of model~(\ref{eq:H_hubbard}) by means of 
a time-dependent variational approach based on the generalized Gutzwiller approximation
 for layered systems~\cite{michele_review,giacomo_phd_prb}. 
 This approach provides a versatile tool for describing, in a non-perturbative way,
the dynamics in the Hubbard model which is governed by 
the  interplay between the hopping terms $t_\parallel, t_\perp$  
and the local Coulomb interaction $U$.

Within this approach,  the effect of the interaction is described in terms of the effective mass renormalization 
$m^*=m/Z$ which is controlled by the interaction $U$ through the quasiparticle weight 
$Z(U)$. In the non-interacting limit $Z(U=0)$=1, whereas at finite interaction $Z<1$ and decreases as a function of $U$.
Eventually, for a critical interaction strength, $U_c$, the system undergoes a metal-to-insulator
Mott transition, corresponding to a vanishing  quasiparticle weight, i.e. $Z(U_c)\rightarrow 0$.
In this regime, quasiparticle excitations are completely suppressed and the dynamics becomes 
dominated by high-energy incoherent excitations at energies $\sim U$~\cite{neq_dmft_review}.
In this work we focus on the thermal dynamics of hot quasi-particles in the correlated metal regime, where Z is finite but significantly smaller than one.

\subsection{Non-equilibrium protocol}
We investigate the thermal dynamics by considering the time evolution, regulated by the interacting Hamiltonian (\ref{eq:H_hubbard}),
of two electronic populations at different temperatures. 
To tackle this non-equilibrium problem we define the following protocol.
We start from the solution of the equilibrium variational problem 
at zero temperature and set an electronic temperature 
on each layer by coupling each layer to an external reservoir 
of electrons with dispersion  $\epsilon_{bath}(\bk)$ and non-zero temperature.
In practice, this is achieved by considering an auxiliary master 
equation for the quasiparticles  and performing a short-time evolution 
of the coupled system until equilibration is reached. 

We first solve the finite temperature equilibration 
for the entire system at a base temperature
that we will refer to as $T_{c0}$, where the subscript "c" stands for "cold" 
and "0" indicates  the instant preceding the impulsive excitation, 
and obtain the finite temperature occupation matrix elements, 
$\langle \cc_{\bk n \sigma }  \ca_{\bk n' \sigma}\rangle (T_{c0})$,
and quasiparticle renormalizations, $Z$.
We then repeat the finite temperature 
equilibration with an higher temperature $T_h > T_{c0}$ 
for a smaller subsystem of five layers.
At time $t=0$ we switch off the coupling with the reservoirs and 
we let the system evolve starting from the condition
\begin{eqnarray}
\langle \cc_{\bk n \sigma }  \ca_{\bk n' \sigma}\rangle (T_h) 
~\quad &\mathrm{for}& \quad n,n'=1,\ldots,5\\
\langle \cc_{\bk n \sigma }  \ca_{\bk n' \sigma}\rangle (T_{c0})~\quad &\mathrm{for}& \quad n,n'=6,\ldots,L.
\end{eqnarray}


\subsection{Observables}
We study the thermal transport by tracking the time evolution of the energy density of the $n^{th}$ layer
\begin{equation}
E_n (t) \equiv \langle \Psi(t) | h_n + \tau_{n,n+1}| \Psi(t) \rangle
\label{eq:ene_layer}
\end{equation}
and the layer-dependent occupation numbers $f_{n}(\bk,t)$, defined as
\begin{equation}
f_{n}(\bk,t) \equiv 
\langle \Psi(t) | \cc_{\bk n \sigma} \ca_{\bk n \sigma} | \Psi(t) \rangle,
\label{eq:nk_layer}
\end{equation}
where in both equations $\left| \Psi(t) \right\rangle$ 
represents the time-evolved Gutzwiller wavefunction.
The observables defined by 
Eqs.~\ref{eq:ene_layer} and~\ref{eq:nk_layer}  are used to extract, respectively, 
the heat flux and the evolution of the local  electronic temperature.

In order to obtain the layer- and time-dependent electronic temperatures, we need to transform the layer-dependent occupation numbers given by Eq. (\ref{eq:nk_layer}) into energy distribution functions by means of a proper variable substitution.
We note that in the equilibrated initial state at time $t=0$ the occupation numbers 
reproduce Fermi-Dirac distribution functions, at the corresponding layer temperatures, expressed as a function of the bath dispersion.
We therefore define the non-equilibrium energy distribution functions by adopting the bath dispersion relation and expressing
the occupation numbers as a function of the energy $\epsilon_{bath}(\bk)$:
\begin{equation}
f_{\mathrm{neq}}^{(n)}(\epsilon,t) \equiv 
f_n(\epsilon_{bath}(\bk),t)
\label{eq:fneq}
\end{equation}
In the rest of the paper we will measure temperatures by setting the bath dispersion equal to the bare electronic dispersion, i.e.
$\epsilon_{bath}(\bk) = \epsilon(\bk)$. 
We mention here that a different choice would corresponds to a simple
rescaling of the base temperature $T_{c0}$. 
In the supplemental (Fig.~S1) 
we show that our results do not depend crucially on this choice.

Typical energy distribution functions are shown in Fig.~\ref{Figure_0} 
for a fixed instant of time at different depths of the layered system.
We find that the non-equilibrium distribution  function $f_{\mathrm{neq}}^{(n)}(\epsilon,t)$ 
can be fitted with a superposition of two equilibrium 
Fermi-Dirac distributions: 
i) a {\it hot} distribution at the temperature $T_h$, fixed by the initial perturbation, and of 
weight $\rho_\mathrm{hot}(n,t)$; ii) a {\it cold} distribution characterized by a 
time- and layer-dependent temperature $T(n,t)$ and of weight $1-\rho_{hot}(n,t)$. 
This decomposition can be written as:
\begin{equation}
f_{\mathrm{neq}}^{(n)}(\epsilon,t) = f_\mathrm{hot}  \rho_\mathrm{hot}(n,t) + f_\mathrm{cold}\left[ 1 - \rho_\mathrm{hot}(n,t) \right],
\label{eq:fneq_def}
\end{equation}
with $0 \leq \rho_\mathrm{hot}(n,t) \leq 1$, $f_\mathrm{hot} = f(\epsilon,T_h)$ and $f_\mathrm{cold} = f(\epsilon,T(n,t))$. Practically, for each fixed $n$ and $t$ values, we fitted $f_{\mathrm{neq}}^{(n)}(\epsilon,t)$, computed via Eq. \ref{eq:fneq}, with the expression given by Eq. \ref{eq:fneq_def}, $\rho_\mathrm{hot}(n,t)$ and $T(n,t)$ being the only two fitting parameters. 

Eq. (\ref{eq:fneq_def}) is found to hold for any instant of time and layer index. This provides
a clear physical interpretation of the transient  propagation of energy and the definition of 
a local time-dependent electronic temperature.
Initially, the perturbation creates a population of hot electrons described by $\rho_\mathrm{hot}(n,t)$
which propagates across the layers.
Remarkably, while the temperature of the hot electrons is fixed at $T_{h}$, 
the temperature of the remaining $1-\rho_\mathrm{hot}(n,t)$ fraction of electrons 
in the "cold" state $T(n,t)$ changes in time
\footnote{This result is confirmed using different fitting procedures 
in which $T_h$ is either considered a fixed parameter or fitting parameter.}. 
We therefore identify $T(n,t)$ as the spatio-temporal evolution
of the local electronic temperature which is determined by the interaction 
between the cold electrons on each layer and the hot electrons
propagating through the system. We pinpoint that $T(n,t)$ 
(referred throughout the manuscript as the "cold" electrons temperature) 
should not be confused with the initial system's temperature $T_{c0}$.

The heat flux $q_n$ at layer $n$ and along the $z$-direction perpendicular 
to the planes is extracted by applying the continuity equation
to the energy density~(\ref{eq:ene_layer})
\begin{equation}
\frac{\partial q_n}{\partial z} + \frac{\partial E_n}{\partial t} = 0
\label{eq:def_heat_flux}
\end{equation}
where the discrete spatial derivative defined with respect to the interlayer 
distance $a$, $\frac{\partial q_n}{\partial z} = (q_{n+1}-q_n)/a$.

\begin{figure*}[t]
\includegraphics[width=\textwidth]{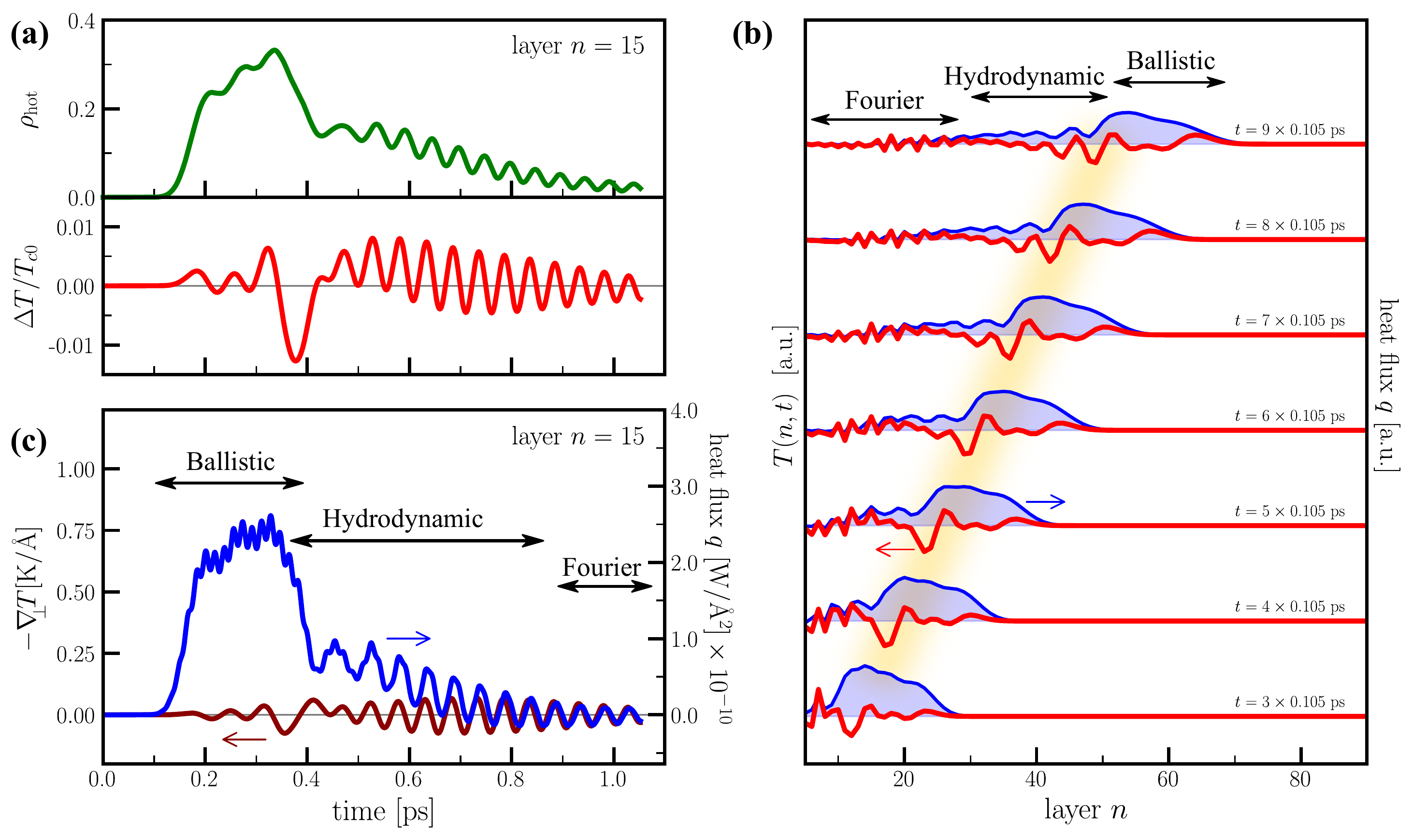}
\caption{\textbf{Sub-picosecond thermal dynamics}.
(a) Dynamics of the hot electron population (top) and relative temperature variation of the "cold" electronic population (bottom) recorded on layer $n=15$.
(b) Layer profiles of the temperature of the "cold" electronic population (red, left axis) 
and of the heat flux (blue, right axis) at different instant of times. The blurred yellow band 
highlights the wave packet of temperature oscillations that follows the 
ballistic front. 
(c) Dynamics of the heat flux (blue, right axis) and temperature gradient of the "cold" electronic population 
(red, left axis) on layer $n=15$.
Arrows indicates the three regimes of thermal transport discussed in the main text.
}
\label{Figure1}
\end{figure*}

\section{Ultrafast thermal dynamics}
In this section we show how this model offers the 
possibility to access different regimes of non-conventional heat transport  
on the sub-picosecond timescale. Each regime will be then  discussed  and analysed 
in the following sections.
We consider $t_{\perp}$=$t_{\parallel}=60~\mathrm{meV}$ and $U = 0.65~\mathrm{eV}$, 
which correspond to an interaction-driven mass renormalization 
$m/m^*\simeq 0.3$, and a lattice spacing $a = 5$~\AA. 
As we shall see, this value of effective mass renormalization 
is consistent with experimental estimates for SrVO$_3$.

Initially the system is at the temperature $T_{c0} \simeq 35~\mathrm{K}$ 
and we create at time $t=0$ a non-equilibrium population of hot 
electrons on the first five layers corresponding to a hot 
temperature $T_h = 10 \times T_{c0}.$

Fig. \ref{Figure1}a reports the results for the time evolution of the hot population weight and the local electronic relative temperature variation, $\Delta T/T_{c0}(n,t)$ 
with $\Delta T=T(n,t)-T_{c0}$, recorded on layer $n$=15, 
which we take as representative of the inner region of the slab.
For times $0<t\lesssim 150~\mathrm{fs}$, both $\rho_\mathrm{hot}(15,t)$ and $T(15,t)$
remain fixed to the equilibrium values $\rho_\mathrm{hot}=0$ and $T=T_{c0}$. 
At $t\sim150~\mathrm{fs}$ the perturbation reaches the $n=15$ layer 
and the dynamics that follows can be neatly divided in three steps.

i) In the time window $150-400~\mathrm{fs}$ the dynamics 
is characterized by a significant increase of $\rho_\mathrm{hot}(15,t)$
highlighting the arrival of the propagating hot electron population.
On this time scale, the electronic relative temperature variation remains 
limited. 
This is indicative of a ballistic regime of energy transport 
in which the energy flows without inducing any heating in the 
underlying quasi-equilibrium distribution.

ii) For $t \gtrsim 400~\mathrm{fs}$ the hot electron population
displays a sharp drop and, concomitantly, we observe the activation
of a fast oscillatory dynamics in the electronic temperature 
of the cold electrons.
Initially the oscillations are centered around a value higher than the initial equilibrium temperature $T_{c0}$ 
indicating that the transit of the ballistic front of hot electrons  induced  
the heating of the population of cold electrons on the layer.

iii) Eventually  the  system equilibrates for $t \gtrsim 0.9~\mathrm{ps}$
with the residual damped temperature oscillations converging to $T_{c0}$.

We gain further insight into the thermal dynamics by comparing the 
dynamics of the local electronic temperature with the heat flux $q_n(t)$ at layer $n$.  
Panel (b) of Fig.~\ref{Figure1} reports the spatial profiles of the 
heat flux (right axis, blue trace) and of the local electronic temperature (left axis, red trace) at fixed instants of time. 
The broad feature at the forefront of the heat flux profile 
indicates the propagation of a ballistic energy front accompanied  
by a small and more localized perturbation of the electronic temperature.
At the back front of the ballistic heat flux, as indicated by the blurred yellow
band in Fig. \ref{Figure1}b, we observe the formation of 
a sharp sinusoidal feature in the spatial profile of the temperature. 
In the time domain, this sharp feature marks the separation between 
the first two dynamical regimes of the local temperature observed in panel (a) 
for the layer $n=15$. 
The presence of this pronounced oscillation of the temperature spatial profile is accompanied by 
weaker temperature oscillations with smaller spatial periodicity in the layers behind the ballistic front.

To fully characterize the thermal dynamics regimes occurring after the 
ballistic front has transited, we further compare  the dynamics of the heat flux with that of the temperature gradient $\nabla_{\perp}T(t) = (T_{n+1} (t) - T_{n}(t) ) / a$ 
perpendicular to the layers.
These quantities are shown in Fig. \ref{Figure1}c for the $n$=15 layer.
In the time window $0.15-0.4~\mathrm{ps}$, the ballistic regime shows up  as a sharp increase of the heat flux 
with no sizeable effect on the temperature gradient. 
In correspondence of the end of the ballistic regime, i.e. the sharp drop of the heat current, an oscillatory dynamics 
is activated for the temperature gradient. The oscillatory dynamics of $\nabla_{\perp}T(t)$ is maintained in the 
0.4-0.9 ps time window, along with a residual positive heat current on the layer. 
At $t\gtrsim~0.9$ ps the heat current displays damped oscillations 
centred around zero indicating the recovery of local 
thermal equilibrium. Remarkably, the equilibration is characterized by the 
synchronization between the dynamics of the temperature gradient 
and the heat flux. In this regime, we can define an instantaneous proportionality between the heat flux and temperature gradient, i.e.  $ q_n (t) \propto -\nabla_\perp T (t)$, indicating
that the heat transfer process is well described by a Fourier-like 
heat transfer law. 

At intermediate times ($0.4<t<0.9~\mathrm{ps}$), before Fourier-like transport sets in, 
there is a residual positive flow of the heat current
with an oscillatory dynamics of $-\nabla_\perp T (t)$ that is not simply proportional to that of $ q_n$($t$). 
This fact reveals the presence
of a new heat transport regime which bridges the ballistic regime 
established at the arrival of the perturbation ($0.15<t< 0.4~\mathrm{ps}$) and the Fourier-like transport setting in at long times after the perturbation has transited ($t>$0.9 ps).
This intermediate regime is characterized by a residual population of hot electrons on the layer and by an oscillatory dynamics of the temperature of the cold electron population.
We  identify this regime as a hydrodynamic transport of heat 
sustained by the exchange of energy between the two sub-populations of hot and cold electrons.
By comparing the dynamics on the single layer (Figs. \ref{Figure1}a,c) with the layer profiles at different 
times (Fig. \ref{Figure1}b), we can observe that the emergence of the hydrodynamic regime coincides with 
transit of the sharp sinusoidal feature in the spatial profile of the temperature at the trailing edge of the heat flux ballistic front. As it will be further discussed in the following sections, this feature can be considered as a {\it{temperature wave-packet}} propagating through the system.

Summarising, the sub-picosecond thermal dynamics of 
electrons displays three subsequent regimes of heat transport:
i) 
the ballistic propagation of energy at the front of the perturbation; 
ii)
the hydrodynamic regime at the trailing edge of the ballistic front. The former is
characterized by a wave-like propagation of the electronic 
temperature; 
iii) a Fourier-like heat transport driving the recovery of thermal equilibrium.
The time and space extension of the three regimes are indicated by 
the arrows in the plots 
of the dynamics  at fixed layer index (see Fig. \ref{Figure1}c) and of spatial profiles 
at fixed time (see Fig. \ref{Figure1}b).
In the remaining of the paper we analyse in detail the different regimes and discuss 
the possibility to control their onset in layered correlated materials.

\section{Ballistic energy transport}
In this section we will address the possibility of controlling the initial ballistic energy 
transport by tuning the microscopic parameters entering in the Hubbard model (\ref{eq:H_hubbard}).
In the ballistic regime, the energy is mostly carried by the population of 
hot electrons at temperature $T_h$.
The energy propagates through hopping 
processes of the hot electrons excited in the first layers. 
Layered correlated materials thus offer two complementary ways 
to control the inter-layer coupling and, in turn, the velocity of propagation 
of the ballistic front, namely tuning either the anisotropy of the system, 
$t_{\perp}$/$t_{\parallel}$, or the strength of the interaction, $U$.
The increase of the latter drives a reduction of the quasiparticle weight $Z$, which
leads to  a larger effective mass for the interlayer motion and a smaller effective hopping, 
$t_\perp^* = Z t_\perp$.

\begin{figure}[t]
\includegraphics[width=\linewidth]{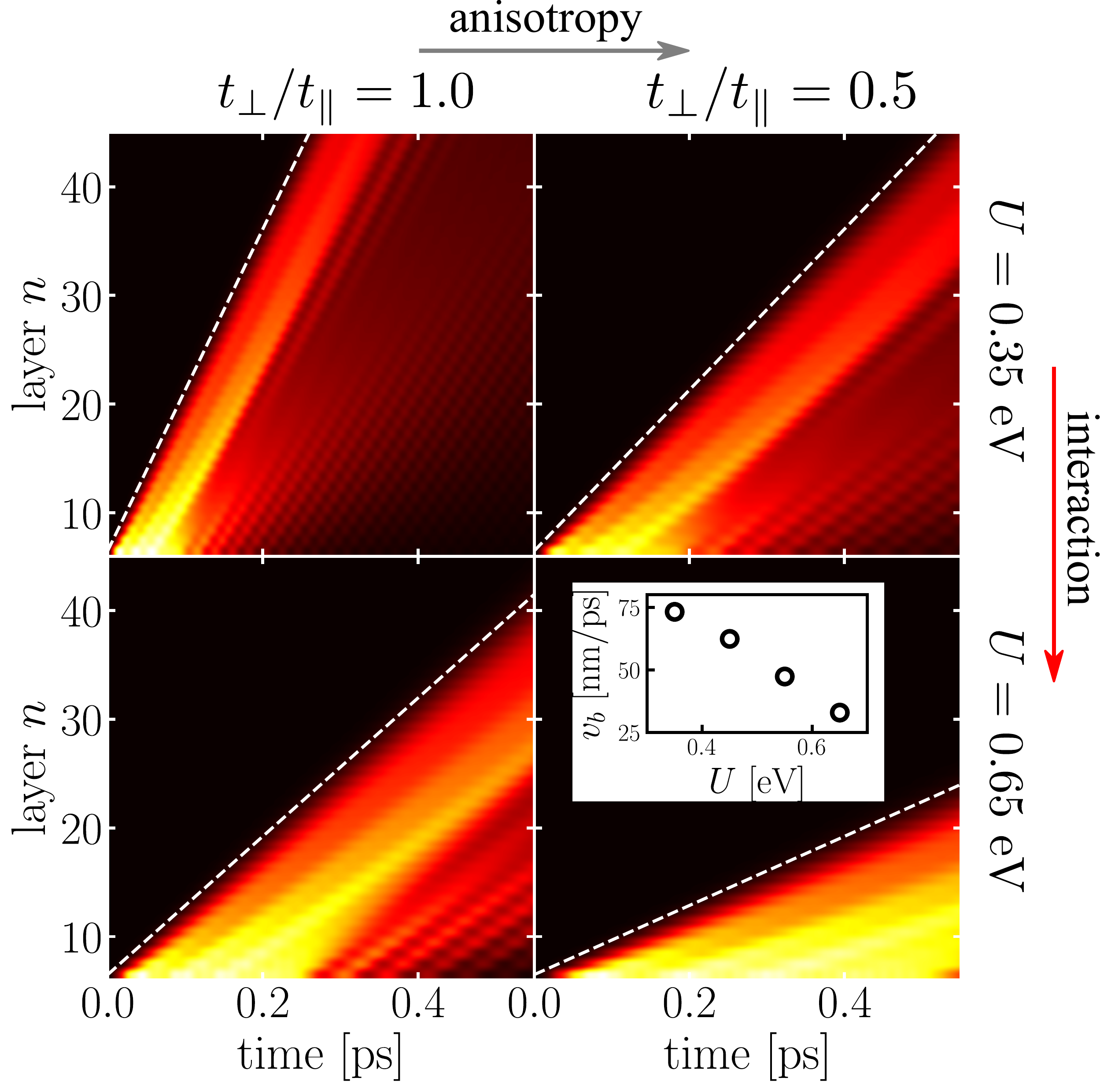}
\caption{\textbf{Control of ballistic energy propagation}. 
The four panels matrix displays the ballistic dynamics of the hot-electron population for varying values of the correlation
strength $U$ and anisotropy $t_\perp/t_\parallel$ ($t_\parallel$=60 meV). The color scale represents the amplitude of $\rho_\mathrm{hot}(n,t)$ (yellow: maximum; black: minimum). 
The inset displays the speed $v_\mathrm{b}$ of the ballistic wavefront (see dashed lines in the main panels) for different values of $U$ at $t_\parallel = t_\perp$.}
\label{Figure_2}
\end{figure}

We show these effects in Fig~\ref{Figure_2} where we report the spatio-temporal 
dynamics of the hot electron population $\rho_\mathrm{hot}(n,t)$ obtained for different values 
of anisotropy (horizontal gray arrow) and relative interaction strength (vertical red arrow). 
Increasing either one the propagation velocity of the wavefront is diminished. 
In the inset we plot the velocity of ballistic propagation 
$v_\mathrm{b}$ as a function of $U$ for $t_{\perp}$/$t_{\parallel}$ = 1. $v_\mathrm{b}$ is defined as the slope of the white dashed line in Fig. \ref{Figure_2}.
The correlation-induced renormalization of $t^*_{\perp}$ strongly suppresses the energy propagation along the $z$-direction. 

For the sake of applications, we note that, in nanosystems with sizes of the order of the ballistic mean free path, the thermal conductivity becomes a size-dependent property \cite{Mingo2005,Munoz2010,Bae2013,Caddeo2017}. 
Nanoengineering of LCM, combined with proper tuning of $U$ and $t_{\perp}$/$t_{\parallel}$, thus offers a new mean to control, on the picosecond timescale, the velocity of ballistic heat pulses and, therefore, the thermal conductivity of nanodevices. 

\section{Hydrodynamic energy transport: emergent electronic temperature waves}
\label{sec:hydrodynamic}

\begin{figure}[t]
\includegraphics[width=\columnwidth]{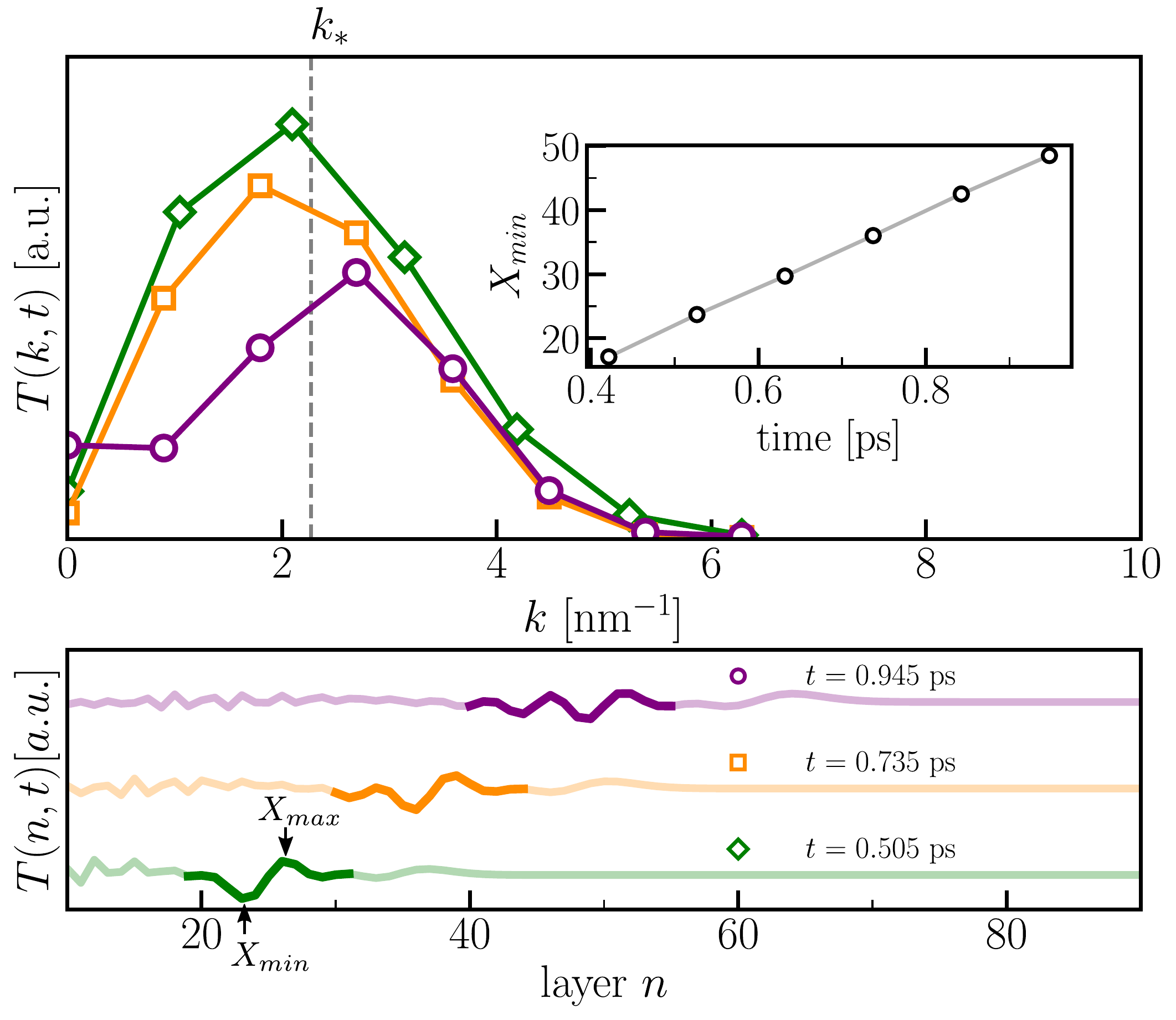}
\caption{\textbf{Spectral analysis in $k$-space of the electronic temperature waves}. 
Top panel:  spatial Fourier transform of the layer profile of the electronic temperature at three 
different times, $t=0.505~\mathrm{ps}$ (diamonds), $t=0.735~\mathrm{ps}$ (squares) and 
$t=0.945~\mathrm{ps}$ (circles). 
The inset shows the position of the minimum of the temperature oscillation, indicated in the bottom panel,
as a function of time.
The vertical dashed line indicated by $k_*$ shows the average of the position of the three peaks.
Bottom panel:
Portions of the temperature profiles at different times used to compute 
the discrete Fourier transform.
}
\label{Figure_3}
\end{figure}

\subsection{Temperature wave-packets}

The results reported in Fig. \ref{Figure1} demonstrate that a purely \textit{electronic}  hydrodynamic transport regime can be achieved in our correlated system on much faster time scales than the
more conventional \textit{phononic} counterpart \cite{Beck1974,Lee2015,Huberman2019}.
Similarly to the phononic case, this hydrodynamic regime manifests itself by a wave-like propagation of temperature oscillations, which emerge after the ballistic front has transited (see arrows in Fig.~\ref{Figure1}c). In this section, we will quantitatively describe the characteristics of temperature wave-like propagation, as it emerges from our microscopic model. 

In order to characterize this regime we track the position of the minimum of the wave packet  $X_\mathrm{min}$
and we observe that it linearly increases in time (See the inset of Fig.~\ref{Figure_3}a), allowing us to estimate the wave packet group velocity from the simple relation $X_\mathrm{min} = v_g t$.
We obtain 
$v_g \sim 30~\mathrm{nm}/\mathrm{ps}$ 
of the same order of magnitude of the ballistic energy wavefront velocity.
A similar result is obtained when tracking the time-dependent maximum of the wave-packet, $X_\mathrm{max}$. 
This result suggests that we can approximately describe the wave packet as a superposition
of weakly-dispersive waves with frequencies $\nu_k = v_g k /2 \pi$. 

In order to identify the barycentric wavevector of the propagating wave packet, in the top panel of Fig. ~\ref{Figure_3} we
report the spatial Fourier transform of the electronic temperature profile 
in the spatial window where the propagating packet is 
present, as highlighted in the three curves of
the bottom panel, which correspond to three different times, $t=0.505~\mathrm{ps}$, $t=0.735~\mathrm{ps}$
and $t=0.945~\mathrm{ps}$.
The small number of layers included in the Fourier window produces spectrally broaden peaks
with the maximum occurring at slightly different $k$ values for different times.
We estimate the peak wavevector by taking the average of the three peaks observed 
at the three chosen times, obtaining $k_* \sim 2.2~\mathrm{nm^{-1}}$, corresponding to a wavelength $\lambda \sim 2.85~\mathrm{nm}$.
Inserting this result in the linear dispersion relation we obtain a frequency $\nu_* \sim 10.5~\mathrm{THz}$.
We notice that in the time domain, and at fixed layer index, this frequency 
corresponds to the inverse of the period of the large amplitude temperature oscillation originating after the transit of the ballistic energy wavefront, as shown in Fig. \ref{Figure1}a.

\subsection{Macroscopic model}
We now compare the predictions of the microscopic model to a phenomenological model
for the description of the hydrodynamic regime characterized by the emergence of 
electronic temperature waves.
This approach recently proved effective in describing phononic temperature wave oscillations in graphite \cite{Gandolfi2019}. 

The phenomenological approach is based on the Dual Phase Lag Model (DPLM)\cite{Tzou2014}, which 
modifies Fourier law by
introducing a causality relation between the onset of $\nabla_{\perp} T$ and the heat flux 
\begin{equation}
q\left(t+\tau_q,z\right)=-\kappa_{T,el}\ \nabla_{\perp} T \left(t+\tau_T,z\right).
\label{DPLM}
\end{equation}
In other words, the DPLM introduces a delay between the time at which the 
temperature gradient $\nabla_{\perp}T$ is established, $t+\tau_{T}$, and the 
time when the interlayer heat flux $q$ sets in, $t+\tau_{q}$. 
The expansion of Eq. \ref{DPLM} to first order, and its combination with the local conservation of energy at time $t$, gives rise to a second order parabolic differential equation for the temperature variation $\Delta T(t,z)$=$T(t,z)-T_{\mathrm{c0}}$. 

We look for wave-like solutions of this differential equation starting from a temperature pulse  
triggered at initial time on the top side of the sample slab. Following Ref.~\citenum{Gandolfi2019},
the pulse can be described by a superposition of plane-waves of real-valued wave vectors $k$
and complex frequencies $\nu$. 
Underdamped plane-wave solutions for $\Delta T(t,z)$ are found if the condition $\tau_q > 2\tau_T$ is met.
These temperature waves are characterized by the complex-valued dispersion relation
\begin{equation}
\nu(k,R,\alpha) = \nu_1(k,R,\alpha) + \mathrm{i} \nu_2(k,R,\alpha),
\end{equation}
where $\nu_{1,2}(k,R,\alpha)$ depend on the wavevector $k$, and on the parameters $R=\frac{\tau_T}{\tau_q}$ 
and thermal diffusivity $\alpha = \frac{\kappa_{T,el}}{C_{el}}$, $\kappa_{T,el}$ and $C_{el}$ 
being the electronic thermal conductivity and specific heat, respectively.

The analytic expressions for the real-valued $\nu_1$ and $\nu_2$, together with the quality factor
defined as $Q(k,R,\alpha) = \frac{\nu_1}{\nu_2}$, are 
reported in Supplementary Information.
In principle, the quantities $R$ and $\alpha$ do depend on the electronic temperature $T$. 
However, since the relative variation $|\Delta T(t)|/T_{c0} \ll 1$ (see Fig.~\ref{Figure1}a), 
the temperature dependence may be taken with respect to the initial base temperature $T_{c0}$, 
i.e.  $R$=$R$($T_{c0}$) and $\alpha$=$\alpha$($T_{c0}).$

In order to reveal under which conditions temperature 
waves are sustained, we exploit the dispersion $\nu_{1}$($k$) and its quality factor $Q$, 
upon insertion of the microscopic parameters relevant for SrVO$_3$ as extracted both from solution of the Hubbard model (previous sections) and supplemented by parameters derived from the literature.
We first determine the quantities $R$ 
 and $\alpha$. 
We identify the time for setting a variation in the temperature gradient,
$\tau_T$, as the electronic thermalization time.
The local thermalization time in SVO is estimated to be as short as $\sim$5 fs 
on the basis of  angle-resolved photoemission spectroscopy \cite{Aizaki2012} and 
optical conductivity \cite{Zhang2015} data (see Supplementary Information). 
We thus set $\tau_{T}$=5 fs. 
This time scale is compatible with the attribution of an instantaneous local temperature on the sub-picosecond time scale, as assumed in the previous sections.
On the other hand, the heat flux dynamics in Fig.~\ref{Figure1}c shows that the
synchronization between $\nabla_{\perp}T$ and $q$ starts at $\sim900~\mathrm{fs}$, 
i.e. $500~\mathrm{fs}$ after the ballistic wavefront has transited through the 15$^{\mathrm{th}}$ layer. 
We can thus assume $\tau_q\simeq$500 fs. 
Based on these assumptions, we obtain $R = \tau_T/\tau_q \sim 0.01$ which is well below the 
threshold $R<0.5$ for the observation of a wavelike behaviour.
While the electronic scattering time is expected to weakly depend on the temperature, the temperature dependence of $\tau_q$ is tested by calculating the solution of the single-band Hubbard model at different base temperatures $T_{c0}$. As shown in Fig. S1, the results demonstrate that $\tau_{q}$ is almost independent of $T_{c0}$, thus allowing to assume a temperature independent value of $R$.
The temperature dependence of the wave frequencies is instead retained through $\alpha$. Specifically, for the case of SVO, $C_{el}$=$\gamma T$ with $\gamma$=2.4$\times10^{2}~\mathrm{Jm}^{-3}\mathrm{K}^{-2}$~\cite{Inoue1998}. As for $\kappa_{T,el}$($T$) we retrieve it from the temperature dependent electrical conductivity, $\sigma(T)$, of SVO single crystals \cite{Inoue1998} upon application of the Wiedemann-Franz-Lorentz relation: $\kappa_{T,el}$=$L\sigma T$, $L$=2.44$\cdot$10$^{-8}$ W$\Omega$K$^{-2}$ being the Lorentz number. 
The temperature-dependent $\kappa_{T,el}$ ranges from $\simeq$10 Wm$^{-1}$K$^{-1}$ at 300 K to $\simeq$20 Wm$^{-1}$K$^{-1}$ at 35 K. 

\begin{figure}[t]
\includegraphics[width=\columnwidth]{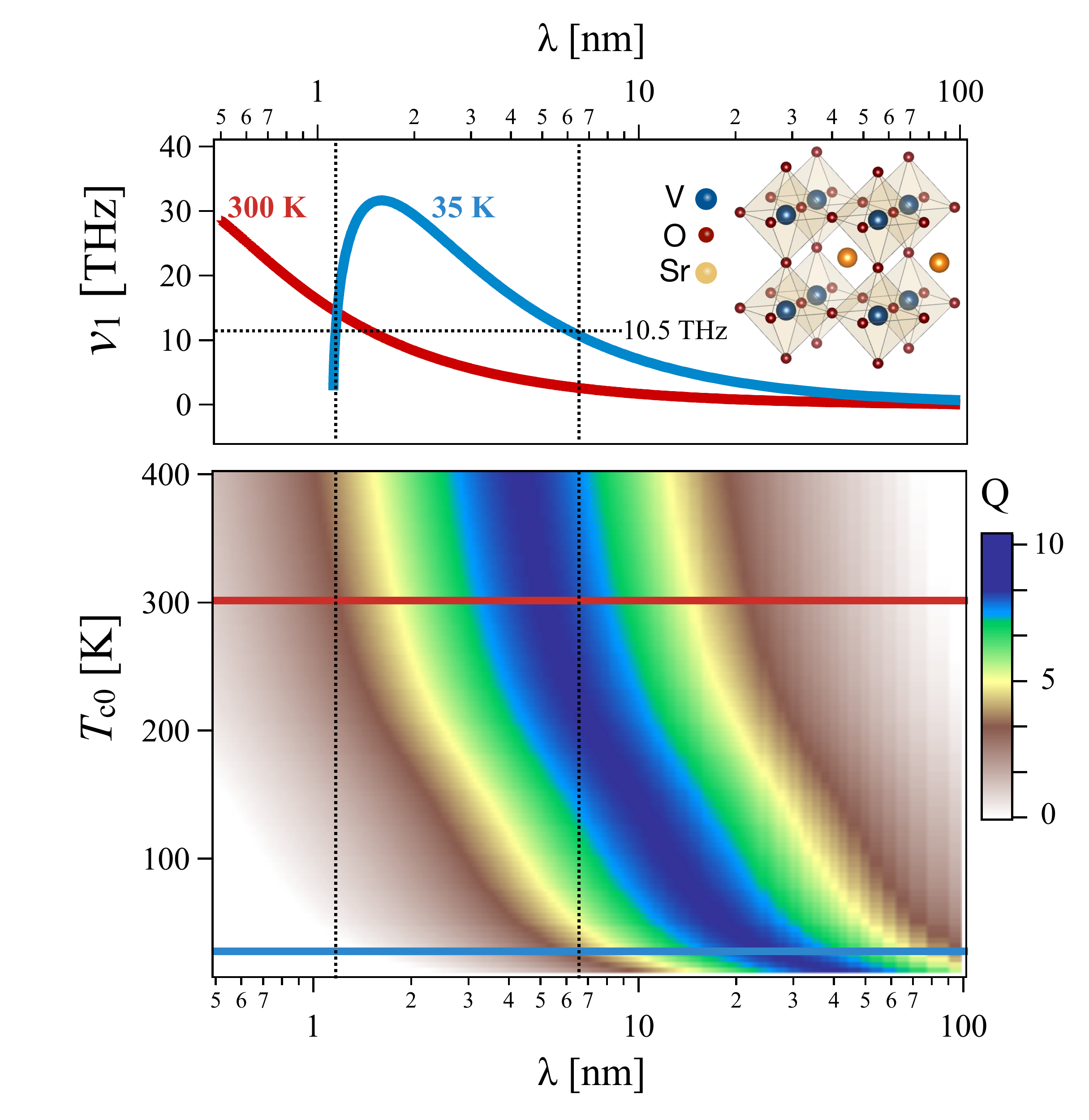}
\caption{\textbf{Temperature wave dispersion in SVO$_3$.} 
Top panel: electronic temperature oscillation frequency, $\nu$=$\omega_1/2\pi$, 
vs oscillation's wavelength, $\lambda$, at a base temperatures $T_{c0}$=35 K (blue line) 
and 300 K (red line). Bottom panel: quality factor (colormap) vs oscillation's wavelength, $\lambda$, and base temperature $T_{c0}$. Calculations based on Ref. \cite{Gandolfi2019} upon insertion of input parameters from experiments, $\alpha$ and $\tau_{T}$, and from non-equilibrium thermal dynamics results from the layered Hubbard model (see text), $\tau_{q}$. In both panels a linear-log plot is adopted.
}
\label{Figure_4}
\end{figure}

With this parameters at hand, in Fig.~\ref{Figure_4} we show the dispersion relation for the temperature oscillation frequency $\nu_1$ (top panel) and the corresponding $Q-$factor (bottom panel) as a function of 
wavelength $\lambda=2\pi/k$ and base temperature $T_{c0}$. The temperature wave frequency $\nu_* \sim 10.5~\mathrm{THz}$, obtained from the microscopic model 
at the base temperature $T_{c0}=35$ K, falls within the range of the allowed frequencies and is compatible with 
two possible 
wavelengths, $\lambda \sim 6.5~\mathrm{nm}$ and $\lambda \sim 1.1~\mathrm{nm}$. These wavelengths correspond 
to $Q-$factors $\sim$ 5 and 0.2, respectively, therefore only  the longest wavelength is expected to be detectable.
This wavelength falls pretty close to the estimate $\lambda \sim 2.85~\mathrm{nm}$ obtained from the microscopic single-band Hubbard model.

Given the quite general assumptions on the parameters of the microscopic model 
and the realistic values used in the phenomenological model, the above comparison shows 
an overall good agreement between the temperature waves dynamics 
obtained from the sub-picoseconds dynamics of the single-band Hubbard model 
and the predictions based on a macroscopic model.
Such an agreement
further confirms that LCM can sustain, in the hydrodynamic regime, temperature waves with 
wavelengths and periods fully compatible with state-of-the-art materials growth techniques 
and time-resolved spectroscopies. 
More in general, the frequencies and $Q$-factor values reported in Fig. \ref{Figure_4} show that the manifestation of temperature waves in LCM can be observed up to temperatures as high as $300~\mathrm{K}$. This is the consequence of the fact that the energy scales controlling the electronic dynamics, i.e. $t_{\parallel}$=$t_{\perp}$=60 meV and $U$=650 meV, correspond to temperatures of $\simeq$ 700 K and $\simeq$ 7000 K respectively. At variance with the phononic case, the sub-picosecond electronic hydrodynamic regime is thus expected to be very robust against temperature, giving rise to the emergence of temperature wave-like oscillations in real materials at ambient conditions.

\subsection{Control of temperature waves in the hydrodynamic regime}
\begin{figure}[t]
\includegraphics[width=0.9\columnwidth]{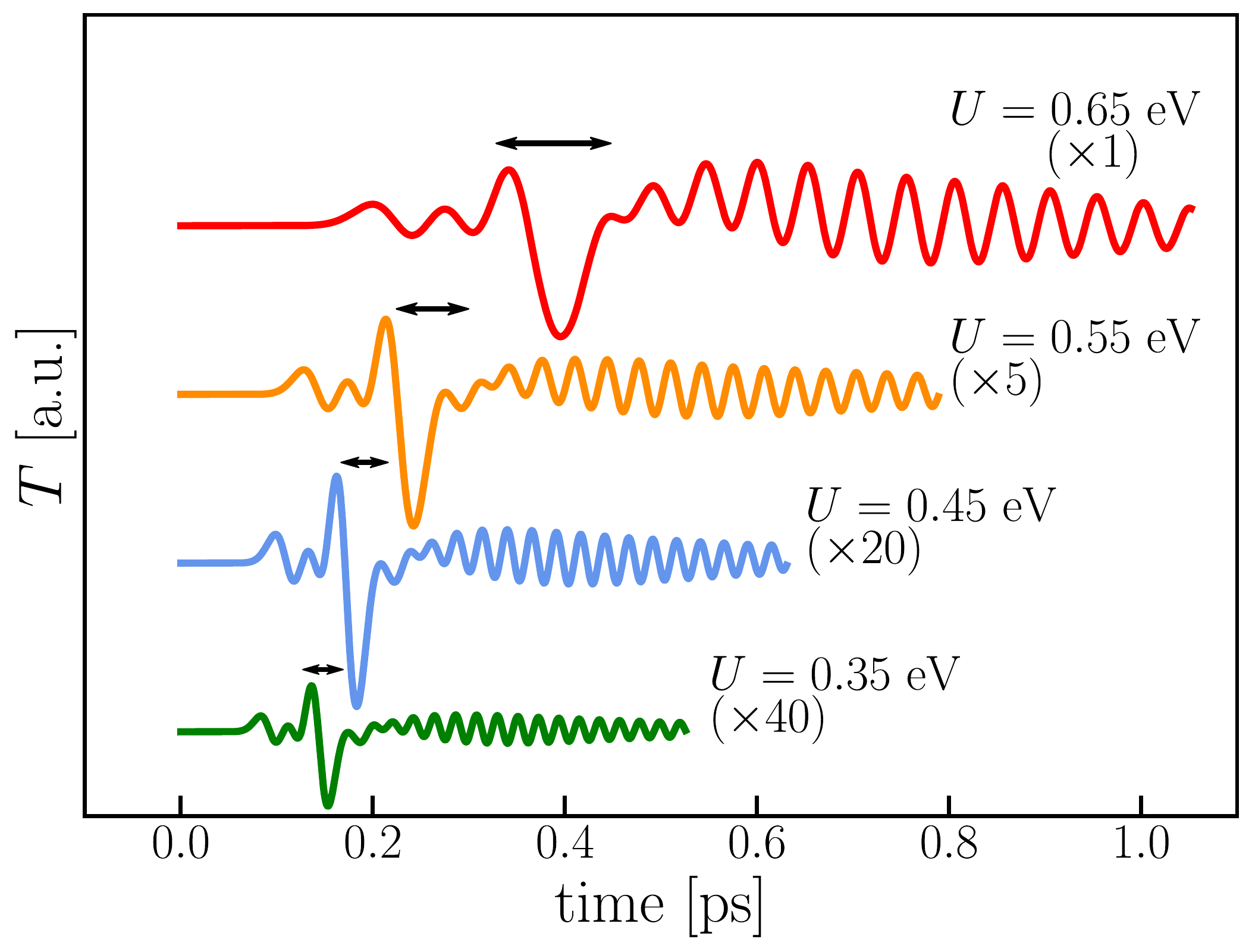}
\caption{\textbf{Control of temperature wave-like oscillations}. 
Temperature $T(t)$ of the "cold" electronic oscillation at the $n=15$
layer for different values of $U$ (the same values used in the inset of Fig.~\ref{Figure_2})
Some of the data have been magnified and the curves shifted for graphical reasons. 
The horizontal arrows highlight the oscillation periods that match 
the frequency  $\nu_*$ extracted from spectral analysis of the
temperature wave packet.
}
\label{Figure_5}
\end{figure}

We end this section by discussing how, similarly to the ballistic transport regime,
the electronic interactions are key to control the wave-like temperature propagation.
In Fig.~\ref{Figure_5} we report the temperature dynamics at layer $n=15$
for different values of the interaction $U$.
We observe that the smaller the interaction the smaller is 
the temperature oscillation amplitude triggered in the population of cold electrons by the transit of the 
hot electron wavefront. 
The data further show that the temperature oscillation periods, indicated in Fig.~\ref{Figure_5} by the black arrows, decrease as the interaction $U$ is decreased. 
In general, the thermal dynamics of quasiparticles become slower as the interaction is increased, a fact also observed for the case of the ballistic energy propagation. This may be traced back to the effect of the correlation-driven renormalization of the quasi-particle effective mass.
Wrapping up, tuning the electronic correlations strength, which controls the quasiparticle effective mass renormalization, can act as a control parameter for the frequency and amplitude of transient temperature waves in LCM. 

\section{Recovery of Fourier-like heat transport}
After the transit of the ballistic heat wavefront and of the temperature wave-packet, the hydrodynamic regime 
gradually evolves into a more conventional dynamics ($t>$0.9 ps in Fig. \ref{Figure1}). 
Here, the non-equilibrium hot electron population has already left the region of interest, 
giving rise to a free oscillatory equilibration dynamics of the temperature of "cold" electrons. 
The wavelength of the temperature oscillation is smaller than that of the temperature wavepacket propagating with speed $v_g$ (hydrodynamic regime), as may be seen in Fig. \ref{Figure1}(b). In the present regime, the oscillation frequency (see Fig. \ref{Figure1}(c)) exactly matches 4$t^*_{\perp}/h$, which 
is the renormalized bandwidth in the direction perpendicular to the layers (Figure \ref{Figure_0}). 
The heat flux left behind by the temperature wavepacket freely oscillates with 
a frequency controlled by $t^*_{\perp}$, which is the only intrinsic energy scale 
of the Hubbard Hamiltonian playing a role on the hundreds femtoseconds timescale. 
The oscillating $q_n$($t$) thus acts as the source for the temperature gradient, which instantaneously follows the temperature variation, i.e. without any delay, as expressed in Fourier law $q\left(t,z\right)=-\kappa_{T,el}\ \nabla_{\perp} T \left(t,z\right)$. For instance, for $U=0.65~\mathrm{eV}$ one has $t^*_{\perp}\simeq 0.33t_{\perp}$=20 meV and the oscillation periods reads $h/4t^*_{\perp} \simeq$ 50 fs. 

Interestingly, we can estimate the electronic thermal conductivity by evaluating the ratio between the \textit{oscillation amplitude} of the heat flux and that of the temperature gradient, i.e. $\kappa_{T,el}$=$\left\lVert q\right\rVert/\left\lVert \nabla_{\perp}T\right\rVert$ where the symbol $\left\lVert ...\right\rVert$ represents the oscillation amplitude. 
As an example of the moderately interacting regime we consider $U=0.45~\mathrm{eV}$, resulting in $\frac{U}{U_c}\sim$0.55, that is quite far from the Mott transition critical point,  $U/U_c \sim$1. In doing so 
we estimate $\kappa_{T,el}\simeq$440 Wm$^{-1}$K$^{-1}$, which is in the order of typical 
zero-frequency electronic thermal conductivity of conventional metals. 
On the other hand, when we increase $U$, the interactions drive a larger temperature gradient (see Fig. S2) which in turn results in a very small value of $\kappa_{T,el}$ \cite{Karrasch2016}. For instance, when $U/U_c$=0.7-0.8, the estimated thermal conductivity is in the range 40-2.5 Wm$^{-1}$K$^{-1}$, a value of the same order of the zero-frequency conductivity reported for SVO \citep{Inoue1998}. Thus, despite its simplicity, the layered Hubbard model predicts the correct order of magnitude of Fourier-like thermal conductivity of materials for a very wide range of correlation strengths.



\section{Conclusions}
In conclusion, we have proposed layered correlated materials as a platform enabling to access a rich variety of heat transport regimes. 
We consider a paradigmatic layered Hubbard model in equilibrium at a given temperature and we impulsively heat one side of the system. The transient dynamics undergoes, on ultrashort space and timescales, a crossover 
between ballistic energy transport and electronic temperature wave-like oscillations in the hydrodynamic regime. 
Eventually, the Fourier-like heat transport regime is recovered on the picosecond time-scale.
Specifically, transition metal oxides thin films and heterostructures, with typical thicknesses and periodicities in the few nanometers range, are here predicted to sustain electronic temperature wave-like oscillations in a parameter space fully compatible with state-of-the-art time resolved calorimetry techniques \cite{Giannetti2016}. 
The temperature oscillation frequency can be tuned via the correlation strength and is predicted to persist up to room temperature. 

The outreach of our results ranges beyond LCM. Among the most interesting applications we foresee is nanoengineering of superlattices made out of correlated materials, allowing for coherent control of temperature waves in nanodevices. For instance, LCM can be grown in heterostructures with control of the physical properties at the level of single atomic layers \cite{Hwang2012} and with the possibility of engineering artificial periodicities to select high-$Q$ modes of temperature waves. The recent introduction of the \textit{temperonic crystal} \cite{Gandolfi2020}, i.e. a periodically modulated structure which behaves like a crystal for temperature waves, provides a new tool to coherently control temperature pulses in correlated heterostructures. Furthermore, strong correlations, and their control via the interlayer twist angle, have recently been reported in graphene superlattices \cite{Cao2018,seyler2019,lu2019}.
The present work paves the way to the control of electronic ballistic propagation and to the engineering of nanodevices exploiting the wavelike nature of the electronic heat transfer on the sub-picosecond timescale.

\section{Acknowledgments}
G.M. acknowledges financial support from the Swiss National Science Foundation through an AMBIZIONE grant.
Part of this work has been supported from the  European Research Council (ERC-319286-QMAC).
M.G. acknowledges financial support from
the CNR Joint Laboratories program 2019-2021.
F.B. acknowledges financial support from Universit\'e de Lyon in the frame of the IDEXLYON Project -Programme Investissements d' Avenir (ANR-16-IDEX-0005) and from Universit\'e Claude Bernard Lyon 1 through the BQR Accueil EC 2019 grant. 
M.G and F.B acknowledge financial support from the MIUR Futuro in ricerca 2013 Grant in the frame of the
ULTRANANO Project (project number: RBFR13NEA4). C.G acknowledge support from Universit\`a Cattolica del Sacro Cuore through D.2.2 and D.3.1 grants. M.C. and C.G. acknowledge financial support from MIUR through the PRIN 2015 (Prot 2015C5SEJJ001) and PRIN 2017 (Prot. 20172H2SC4$\_$005) programs.

\bibliography{biblio_temperonics}

\providecommand{\noopsort}[1]{}\providecommand{\singleletter}[1]{#1}%
\begin{thebibliography}{54}%
\makeatletter
\providecommand \@ifxundefined [1]{%
 \@ifx{#1\undefined}
}%
\providecommand \@ifnum [1]{%
 \ifnum #1\expandafter \@firstoftwo
 \else \expandafter \@secondoftwo
 \fi
}%
\providecommand \@ifx [1]{%
 \ifx #1\expandafter \@firstoftwo
 \else \expandafter \@secondoftwo
 \fi
}%
\providecommand \natexlab [1]{#1}%
\providecommand \enquote  [1]{``#1''}%
\providecommand \bibnamefont  [1]{#1}%
\providecommand \bibfnamefont [1]{#1}%
\providecommand \citenamefont [1]{#1}%
\providecommand \href@noop [0]{\@secondoftwo}%
\providecommand \href [0]{\begingroup \@sanitize@url \@href}%
\providecommand \@href[1]{\@@startlink{#1}\@@href}%
\providecommand \@@href[1]{\endgroup#1\@@endlink}%
\providecommand \@sanitize@url [0]{\catcode `\\12\catcode `\$12\catcode
  `\&12\catcode `\#12\catcode `\^12\catcode `\_12\catcode `\%12\relax}%
\providecommand \@@startlink[1]{}%
\providecommand \@@endlink[0]{}%
\providecommand \url  [0]{\begingroup\@sanitize@url \@url }%
\providecommand \@url [1]{\endgroup\@href {#1}{\urlprefix }}%
\providecommand \urlprefix  [0]{URL }%
\providecommand \Eprint [0]{\href }%
\providecommand \doibase [0]{http://dx.doi.org/}%
\providecommand \selectlanguage [0]{\@gobble}%
\providecommand \bibinfo  [0]{\@secondoftwo}%
\providecommand \bibfield  [0]{\@secondoftwo}%
\providecommand \translation [1]{[#1]}%
\providecommand \BibitemOpen [0]{}%
\providecommand \bibitemStop [0]{}%
\providecommand \bibitemNoStop [0]{.\EOS\space}%
\providecommand \EOS [0]{\spacefactor3000\relax}%
\providecommand \BibitemShut  [1]{\csname bibitem#1\endcsname}%
\let\auto@bib@innerbib\@empty
\bibitem [{\citenamefont {Chen}(2005)}]{Chen2005}%
  \BibitemOpen
  \bibfield  {author} {\bibinfo {author} {\bibfnamefont {Gang}\ \bibnamefont
  {Chen}},\ }\href@noop {} {\emph {\bibinfo {title} {Nanoscale energy transport
  and conversion}}}\ (\bibinfo  {publisher} {Oxford University Press},\
  \bibinfo {year} {2005})\BibitemShut {NoStop}%
\bibitem [{\citenamefont {Volz}\ \emph {et~al.}(2016)\citenamefont {Volz},
  \citenamefont {Ordonez-Miranda}, \citenamefont {Shchepetov}, \citenamefont
  {Prunnila}, \citenamefont {Ahopelto}, \citenamefont {Pezeril}, \citenamefont
  {Vaudel}, \citenamefont {Gusev}, \citenamefont {Ruello}, \citenamefont {Weig}
  \emph {et~al.}}]{volz2016}%
  \BibitemOpen
  \bibfield  {author} {\bibinfo {author} {\bibfnamefont {Sebastian}\
  \bibnamefont {Volz}}, \bibinfo {author} {\bibfnamefont {Jose}\ \bibnamefont
  {Ordonez-Miranda}}, \bibinfo {author} {\bibfnamefont {Andrey}\ \bibnamefont
  {Shchepetov}}, \bibinfo {author} {\bibfnamefont {Mika}\ \bibnamefont
  {Prunnila}}, \bibinfo {author} {\bibfnamefont {Jouni}\ \bibnamefont
  {Ahopelto}}, \bibinfo {author} {\bibfnamefont {Thomas}\ \bibnamefont
  {Pezeril}}, \bibinfo {author} {\bibfnamefont {Gwenaelle}\ \bibnamefont
  {Vaudel}}, \bibinfo {author} {\bibfnamefont {Vitaly}\ \bibnamefont {Gusev}},
  \bibinfo {author} {\bibfnamefont {Pascal}\ \bibnamefont {Ruello}}, \bibinfo
  {author} {\bibfnamefont {Eva~M}\ \bibnamefont {Weig}},  \emph {et~al.},\
  }\bibfield  {title} {\enquote {\bibinfo {title} {Nanophononics: state of the
  art and perspectives},}\ }\href@noop {} {\bibfield  {journal} {\bibinfo
  {journal} {The European Physical Journal B}\ }\textbf {\bibinfo {volume}
  {89}},\ \bibinfo {pages} {15} (\bibinfo {year} {2016})}\BibitemShut {NoStop}%
\bibitem [{\citenamefont {Li}\ \emph {et~al.}(2012)\citenamefont {Li},
  \citenamefont {Ren}, \citenamefont {Wang}, \citenamefont {Zhang},
  \citenamefont {H\"anggi},\ and\ \citenamefont {Li}}]{li2012}%
  \BibitemOpen
  \bibfield  {author} {\bibinfo {author} {\bibfnamefont {Nianbei}\ \bibnamefont
  {Li}}, \bibinfo {author} {\bibfnamefont {Jie}\ \bibnamefont {Ren}}, \bibinfo
  {author} {\bibfnamefont {Lei}\ \bibnamefont {Wang}}, \bibinfo {author}
  {\bibfnamefont {Gang}\ \bibnamefont {Zhang}}, \bibinfo {author}
  {\bibfnamefont {Peter}\ \bibnamefont {H\"anggi}}, \ and\ \bibinfo {author}
  {\bibfnamefont {Baowen}\ \bibnamefont {Li}},\ }\bibfield  {title} {\enquote
  {\bibinfo {title} {Colloquium: Phononics: Manipulating heat flow with
  electronic analogs and beyond},}\ }\href {\doibase
  10.1103/RevModPhys.84.1045} {\bibfield  {journal} {\bibinfo  {journal} {Rev.
  Mod. Phys.}\ }\textbf {\bibinfo {volume} {84}},\ \bibinfo {pages}
  {1045--1066} (\bibinfo {year} {2012})}\BibitemShut {NoStop}%
\bibitem [{\citenamefont {Luo}\ and\ \citenamefont {Chen}(2013)}]{luo2013}%
  \BibitemOpen
  \bibfield  {author} {\bibinfo {author} {\bibfnamefont {Tengfei}\ \bibnamefont
  {Luo}}\ and\ \bibinfo {author} {\bibfnamefont {Gang}\ \bibnamefont {Chen}},\
  }\bibfield  {title} {\enquote {\bibinfo {title} {Nanoscale heat transfer -
  from computation to experiment},}\ }\href {\doibase 10.1039/C2CP43771F}
  {\bibfield  {journal} {\bibinfo  {journal} {Phys. Chem. Chem. Phys.}\
  }\textbf {\bibinfo {volume} {15}},\ \bibinfo {pages} {3389--3412} (\bibinfo
  {year} {2013})}\BibitemShut {NoStop}%
\bibitem [{\citenamefont {Cahill}\ \emph {et~al.}(2014)\citenamefont {Cahill},
  \citenamefont {Braun}, \citenamefont {Chen}, \citenamefont {Clarke},
  \citenamefont {Fan}, \citenamefont {Goodson}, \citenamefont {Keblinski},
  \citenamefont {King}, \citenamefont {Mahan}, \citenamefont {Majumdar} \emph
  {et~al.}}]{cahill2014nanoscale}%
  \BibitemOpen
  \bibfield  {author} {\bibinfo {author} {\bibfnamefont {David~G}\ \bibnamefont
  {Cahill}}, \bibinfo {author} {\bibfnamefont {Paul~V}\ \bibnamefont {Braun}},
  \bibinfo {author} {\bibfnamefont {Gang}\ \bibnamefont {Chen}}, \bibinfo
  {author} {\bibfnamefont {David~R}\ \bibnamefont {Clarke}}, \bibinfo {author}
  {\bibfnamefont {Shanhui}\ \bibnamefont {Fan}}, \bibinfo {author}
  {\bibfnamefont {Kenneth~E}\ \bibnamefont {Goodson}}, \bibinfo {author}
  {\bibfnamefont {Pawel}\ \bibnamefont {Keblinski}}, \bibinfo {author}
  {\bibfnamefont {William~P}\ \bibnamefont {King}}, \bibinfo {author}
  {\bibfnamefont {Gerald~D}\ \bibnamefont {Mahan}}, \bibinfo {author}
  {\bibfnamefont {Arun}\ \bibnamefont {Majumdar}},  \emph {et~al.},\ }\bibfield
   {title} {\enquote {\bibinfo {title} {Nanoscale thermal transport. ii.
  2003--2012},}\ }\href@noop {} {\bibfield  {journal} {\bibinfo  {journal}
  {Appl. Phys. Rev.}\ }\textbf {\bibinfo {volume} {1}},\ \bibinfo {pages}
  {011305} (\bibinfo {year} {2014})}\BibitemShut {NoStop}%
\bibitem [{\citenamefont {Siemens}\ \emph {et~al.}(2010)\citenamefont
  {Siemens}, \citenamefont {Li}, \citenamefont {Yang}, \citenamefont {Nelson},
  \citenamefont {Anderson}, \citenamefont {Murnane},\ and\ \citenamefont
  {Kapteyn}}]{Siemens2010}%
  \BibitemOpen
  \bibfield  {author} {\bibinfo {author} {\bibfnamefont {Mark~E.}\ \bibnamefont
  {Siemens}}, \bibinfo {author} {\bibfnamefont {Qing}\ \bibnamefont {Li}},
  \bibinfo {author} {\bibfnamefont {Ronggui}\ \bibnamefont {Yang}}, \bibinfo
  {author} {\bibfnamefont {Keith~A.}\ \bibnamefont {Nelson}}, \bibinfo {author}
  {\bibfnamefont {Erik~H.}\ \bibnamefont {Anderson}}, \bibinfo {author}
  {\bibfnamefont {Margaret~M.}\ \bibnamefont {Murnane}}, \ and\ \bibinfo
  {author} {\bibfnamefont {Henry~C.}\ \bibnamefont {Kapteyn}},\ }\bibfield
  {title} {\enquote {\bibinfo {title} {Quasi-ballistic thermal transport from
  nanoscale interfaces observed using ultrafast coherent soft x-ray beams},}\
  }\href@noop {} {\bibfield  {journal} {\bibinfo  {journal} {Nat. Mater.}\
  }\textbf {\bibinfo {volume} {9}},\ \bibinfo {pages} {26} (\bibinfo {year}
  {2010})}\BibitemShut {NoStop}%
\bibitem [{\citenamefont {Minnich}\ \emph {et~al.}(2011)\citenamefont
  {Minnich}, \citenamefont {Johnson}, \citenamefont {Schmidt}, \citenamefont
  {Esfarjani}, \citenamefont {Dresselhaus}, \citenamefont {Nelson},\ and\
  \citenamefont {Chen}}]{Minnich2011}%
  \BibitemOpen
  \bibfield  {author} {\bibinfo {author} {\bibfnamefont {J.}~\bibnamefont
  {Minnich}}, \bibinfo {author} {\bibfnamefont {J.~A.}\ \bibnamefont
  {Johnson}}, \bibinfo {author} {\bibfnamefont {A.~J.}\ \bibnamefont
  {Schmidt}}, \bibinfo {author} {\bibfnamefont {K.}~\bibnamefont {Esfarjani}},
  \bibinfo {author} {\bibfnamefont {M.~S.}\ \bibnamefont {Dresselhaus}},
  \bibinfo {author} {\bibfnamefont {K.~A.}\ \bibnamefont {Nelson}}, \ and\
  \bibinfo {author} {\bibfnamefont {G.}~\bibnamefont {Chen}},\ }\bibfield
  {title} {\enquote {\bibinfo {title} {Thermal conductivity spectroscopy
  technique to measure phonon mean free paths},}\ }\href@noop {} {\bibfield
  {journal} {\bibinfo  {journal} {Phys. Rev. Lett.}\ }\textbf {\bibinfo
  {volume} {107}},\ \bibinfo {pages} {095901} (\bibinfo {year}
  {2011})}\BibitemShut {NoStop}%
\bibitem [{\citenamefont {Johnson}\ \emph {et~al.}(2013)\citenamefont
  {Johnson}, \citenamefont {Maznev}, \citenamefont {Cuffe}, \citenamefont
  {Eliason}, \citenamefont {Minnich}, \citenamefont {Kehoe}, \citenamefont
  {Torres}, \citenamefont {Chen},\ and\ \citenamefont {Nelson}}]{Johnson2013}%
  \BibitemOpen
  \bibfield  {author} {\bibinfo {author} {\bibfnamefont {Jeremy~A.}\
  \bibnamefont {Johnson}}, \bibinfo {author} {\bibfnamefont {A.~A.}\
  \bibnamefont {Maznev}}, \bibinfo {author} {\bibfnamefont {John}\ \bibnamefont
  {Cuffe}}, \bibinfo {author} {\bibfnamefont {Jeffrey~K.}\ \bibnamefont
  {Eliason}}, \bibinfo {author} {\bibfnamefont {Austin~J.}\ \bibnamefont
  {Minnich}}, \bibinfo {author} {\bibfnamefont {Timothy}\ \bibnamefont
  {Kehoe}}, \bibinfo {author} {\bibfnamefont {Clivia M.~Sotomayor}\
  \bibnamefont {Torres}}, \bibinfo {author} {\bibfnamefont {Gang}\ \bibnamefont
  {Chen}}, \ and\ \bibinfo {author} {\bibfnamefont {Keith~A.}\ \bibnamefont
  {Nelson}},\ }\bibfield  {title} {\enquote {\bibinfo {title} {Direct
  measurement of room-temperature nondiffusive thermal transport over micron
  distances in a silicon membrane},}\ }\href {\doibase
  10.1103/PhysRevLett.110.025901} {\bibfield  {journal} {\bibinfo  {journal}
  {Phys. Rev. Lett.}\ }\textbf {\bibinfo {volume} {110}},\ \bibinfo {pages}
  {025901} (\bibinfo {year} {2013})}\BibitemShut {NoStop}%
\bibitem [{\citenamefont {Hoogeboom-Pot}\ \emph {et~al.}(2015)\citenamefont
  {Hoogeboom-Pot}, \citenamefont {Hernandez-Charpak}, \citenamefont {Gu},
  \citenamefont {Frazer}, \citenamefont {Anderson}, \citenamefont {Chao},
  \citenamefont {Falcone}, \citenamefont {Yang}, \citenamefont {Murnane},
  \citenamefont {Kapteyn},\ and\ \citenamefont {Nardi}}]{Hoogeboom2015}%
  \BibitemOpen
  \bibfield  {author} {\bibinfo {author} {\bibfnamefont {Kathleen~M.}\
  \bibnamefont {Hoogeboom-Pot}}, \bibinfo {author} {\bibfnamefont {Jorge~N.}\
  \bibnamefont {Hernandez-Charpak}}, \bibinfo {author} {\bibfnamefont
  {Xiaokun}\ \bibnamefont {Gu}}, \bibinfo {author} {\bibfnamefont {Travis~D.}\
  \bibnamefont {Frazer}}, \bibinfo {author} {\bibfnamefont {Erik~H.}\
  \bibnamefont {Anderson}}, \bibinfo {author} {\bibfnamefont {Weilun}\
  \bibnamefont {Chao}}, \bibinfo {author} {\bibfnamefont {Roger~W.}\
  \bibnamefont {Falcone}}, \bibinfo {author} {\bibfnamefont {Ronggui}\
  \bibnamefont {Yang}}, \bibinfo {author} {\bibfnamefont {Margaret~M.}\
  \bibnamefont {Murnane}}, \bibinfo {author} {\bibfnamefont {Henry~C.}\
  \bibnamefont {Kapteyn}}, \ and\ \bibinfo {author} {\bibfnamefont {Damiano}\
  \bibnamefont {Nardi}},\ }\bibfield  {title} {\enquote {\bibinfo {title} {A
  new regime of nanoscale thermal transport: Collective diffusion increases
  dissipation efficiency},}\ }\href@noop {} {\bibfield  {journal} {\bibinfo
  {journal} {Proc. Natl. Acad. Sci. USA}\ }\textbf {\bibinfo {volume} {112}},\
  \bibinfo {pages} {4846} (\bibinfo {year} {2015})}\BibitemShut {NoStop}%
\bibitem [{\citenamefont {Chen}\ \emph {et~al.}(2018)\citenamefont {Chen},
  \citenamefont {Hua}, \citenamefont {Zhang}, \citenamefont {Ravichandran},\
  and\ \citenamefont {Minnich}}]{chen2018}%
  \BibitemOpen
  \bibfield  {author} {\bibinfo {author} {\bibfnamefont {Xiangwen}\
  \bibnamefont {Chen}}, \bibinfo {author} {\bibfnamefont {Chengyun}\
  \bibnamefont {Hua}}, \bibinfo {author} {\bibfnamefont {Hang}\ \bibnamefont
  {Zhang}}, \bibinfo {author} {\bibfnamefont {Navaneetha~K.}\ \bibnamefont
  {Ravichandran}}, \ and\ \bibinfo {author} {\bibfnamefont {Austin~J.}\
  \bibnamefont {Minnich}},\ }\bibfield  {title} {\enquote {\bibinfo {title}
  {Quasiballistic thermal transport from nanoscale heaters and the role of the
  spatial frequency},}\ }\href {\doibase 10.1103/PhysRevApplied.10.054068}
  {\bibfield  {journal} {\bibinfo  {journal} {Phys. Rev. Applied}\ }\textbf
  {\bibinfo {volume} {10}},\ \bibinfo {pages} {054068} (\bibinfo {year}
  {2018})}\BibitemShut {NoStop}%
\bibitem [{\citenamefont {Frazer}\ \emph {et~al.}(2019)\citenamefont {Frazer},
  \citenamefont {Knobloch}, \citenamefont {Hoogeboom-Pot}, \citenamefont
  {Nardi}, \citenamefont {Chao}, \citenamefont {Falcone}, \citenamefont
  {Murnane}, \citenamefont {Kapteyn},\ and\ \citenamefont
  {Hernandez-Charpak}}]{Frazer2019}%
  \BibitemOpen
  \bibfield  {author} {\bibinfo {author} {\bibfnamefont {Travis~D.}\
  \bibnamefont {Frazer}}, \bibinfo {author} {\bibfnamefont {Joshua~L.}\
  \bibnamefont {Knobloch}}, \bibinfo {author} {\bibfnamefont {Kathleen~M.}\
  \bibnamefont {Hoogeboom-Pot}}, \bibinfo {author} {\bibfnamefont {Damiano}\
  \bibnamefont {Nardi}}, \bibinfo {author} {\bibfnamefont {Weilun}\
  \bibnamefont {Chao}}, \bibinfo {author} {\bibfnamefont {Roger~W.}\
  \bibnamefont {Falcone}}, \bibinfo {author} {\bibfnamefont {Margaret~M.}\
  \bibnamefont {Murnane}}, \bibinfo {author} {\bibfnamefont {Henry~C.}\
  \bibnamefont {Kapteyn}}, \ and\ \bibinfo {author} {\bibfnamefont {Jorge~N.}\
  \bibnamefont {Hernandez-Charpak}},\ }\bibfield  {title} {\enquote {\bibinfo
  {title} {Engineering nanoscale thermal transport: Size- and spacing-dependent
  cooling of nanostructures},}\ }\href {\doibase
  10.1103/PhysRevApplied.11.024042} {\bibfield  {journal} {\bibinfo  {journal}
  {Phys. Rev. Applied}\ }\textbf {\bibinfo {volume} {11}},\ \bibinfo {pages}
  {024042} (\bibinfo {year} {2019})}\BibitemShut {NoStop}%
\bibitem [{\citenamefont {Guyer}\ and\ \citenamefont
  {Krumhansl}(1966)}]{Guyer1966}%
  \BibitemOpen
  \bibfield  {author} {\bibinfo {author} {\bibfnamefont {R.~A.}\ \bibnamefont
  {Guyer}}\ and\ \bibinfo {author} {\bibfnamefont {J.~A.}\ \bibnamefont
  {Krumhansl}},\ }\bibfield  {title} {\enquote {\bibinfo {title} {Thermal
  conductivity, second sound, and phonon hydrodynamic phenomena in nonmetallic
  crystals},}\ }\href@noop {} {\bibfield  {journal} {\bibinfo  {journal} {Phys.
  Rev.}\ }\textbf {\bibinfo {volume} {148}},\ \bibinfo {pages} {778--788}
  (\bibinfo {year} {1966})}\BibitemShut {NoStop}%
\bibitem [{\citenamefont {Beck}\ \emph {et~al.}(1974)\citenamefont {Beck},
  \citenamefont {Meier},\ and\ \citenamefont {Thellun}}]{Beck1974}%
  \BibitemOpen
  \bibfield  {author} {\bibinfo {author} {\bibfnamefont {H.}~\bibnamefont
  {Beck}}, \bibinfo {author} {\bibfnamefont {P.F.}\ \bibnamefont {Meier}}, \
  and\ \bibinfo {author} {\bibfnamefont {A.}~\bibnamefont {Thellun}},\
  }\bibfield  {title} {\enquote {\bibinfo {title} {Phonon hydrodynamics in
  solids},}\ }\href@noop {} {\bibfield  {journal} {\bibinfo  {journal} {Phys.
  Stat. Sol. (a)}\ }\textbf {\bibinfo {volume} {24}},\ \bibinfo {pages}
  {11--63} (\bibinfo {year} {1974})}\BibitemShut {NoStop}%
\bibitem [{\citenamefont {Lee}\ \emph {et~al.}(2015)\citenamefont {Lee},
  \citenamefont {Broido}, \citenamefont {Esfarjani},\ and\ \citenamefont
  {Chen}}]{Lee2015}%
  \BibitemOpen
  \bibfield  {author} {\bibinfo {author} {\bibfnamefont {Sangyeop}\
  \bibnamefont {Lee}}, \bibinfo {author} {\bibfnamefont {David}\ \bibnamefont
  {Broido}}, \bibinfo {author} {\bibfnamefont {Keivan}\ \bibnamefont
  {Esfarjani}}, \ and\ \bibinfo {author} {\bibfnamefont {Gang}\ \bibnamefont
  {Chen}},\ }\bibfield  {title} {\enquote {\bibinfo {title} {{Hydrodynamic
  phonon transport in suspended graphene}},}\ }\href@noop {} {\bibfield
  {journal} {\bibinfo  {journal} {Nature Communications}\ }\textbf {\bibinfo
  {volume} {6}},\ \bibinfo {pages} {6290} (\bibinfo {year} {2015})}\BibitemShut
  {NoStop}%
\bibitem [{\citenamefont {Ding}\ \emph {et~al.}(2018)\citenamefont {Ding},
  \citenamefont {Zhou}, \citenamefont {Song}, \citenamefont {Chiloyan},
  \citenamefont {Li}, \citenamefont {Liu},\ and\ \citenamefont
  {Chen}}]{ding2018}%
  \BibitemOpen
  \bibfield  {author} {\bibinfo {author} {\bibfnamefont {Zhiwei}\ \bibnamefont
  {Ding}}, \bibinfo {author} {\bibfnamefont {Jiawei}\ \bibnamefont {Zhou}},
  \bibinfo {author} {\bibfnamefont {Bai}\ \bibnamefont {Song}}, \bibinfo
  {author} {\bibfnamefont {Vazrik}\ \bibnamefont {Chiloyan}}, \bibinfo {author}
  {\bibfnamefont {Mingda}\ \bibnamefont {Li}}, \bibinfo {author} {\bibfnamefont
  {Te-Huan}\ \bibnamefont {Liu}}, \ and\ \bibinfo {author} {\bibfnamefont
  {Gang}\ \bibnamefont {Chen}},\ }\bibfield  {title} {\enquote {\bibinfo
  {title} {Phonon hydrodynamic heat conduction and knudsen minimum in
  graphite},}\ }\href {\doibase 10.1021/acs.nanolett.7b04932} {\bibfield
  {journal} {\bibinfo  {journal} {Nano Letters}\ }\textbf {\bibinfo {volume}
  {18}},\ \bibinfo {pages} {638--649} (\bibinfo {year} {2018})}\BibitemShut
  {NoStop}%
\bibitem [{\citenamefont {Cepellotti}\ \emph {et~al.}(2015)\citenamefont
  {Cepellotti}, \citenamefont {Fugallo}, \citenamefont {Paulatto},
  \citenamefont {Lazzeri}, \citenamefont {Mauri},\ and\ \citenamefont
  {Marzari}}]{cepellotti2015}%
  \BibitemOpen
  \bibfield  {author} {\bibinfo {author} {\bibfnamefont {Andrea}\ \bibnamefont
  {Cepellotti}}, \bibinfo {author} {\bibfnamefont {Giorgia}\ \bibnamefont
  {Fugallo}}, \bibinfo {author} {\bibfnamefont {Lorenzo}\ \bibnamefont
  {Paulatto}}, \bibinfo {author} {\bibfnamefont {Michele}\ \bibnamefont
  {Lazzeri}}, \bibinfo {author} {\bibfnamefont {Francesco}\ \bibnamefont
  {Mauri}}, \ and\ \bibinfo {author} {\bibfnamefont {Nicola}\ \bibnamefont
  {Marzari}},\ }\bibfield  {title} {\enquote {\bibinfo {title} {Phonon
  hydrodynamics in two-dimensional materials},}\ }\href@noop {} {\bibfield
  {journal} {\bibinfo  {journal} {Nature communications}\ }\textbf {\bibinfo
  {volume} {6}},\ \bibinfo {pages} {1--7} (\bibinfo {year} {2015})}\BibitemShut
  {NoStop}%
\bibitem [{\citenamefont {Li}\ and\ \citenamefont {Lee}(2019)}]{Li2019_PRB}%
  \BibitemOpen
  \bibfield  {author} {\bibinfo {author} {\bibfnamefont {Xun}\ \bibnamefont
  {Li}}\ and\ \bibinfo {author} {\bibfnamefont {Sangyeop}\ \bibnamefont
  {Lee}},\ }\bibfield  {title} {\enquote {\bibinfo {title} {Crossover of
  ballistic, hydrodynamic, and diffusive phonon transport in suspended
  graphene},}\ }\href {\doibase 10.1103/PhysRevB.99.085202} {\bibfield
  {journal} {\bibinfo  {journal} {Phys. Rev. B}\ }\textbf {\bibinfo {volume}
  {99}},\ \bibinfo {pages} {085202} (\bibinfo {year} {2019})}\BibitemShut
  {NoStop}%
\bibitem [{\citenamefont {Gandolfi}\ \emph {et~al.}(2019)\citenamefont
  {Gandolfi}, \citenamefont {Benetti}, \citenamefont {Glorieux}, \citenamefont
  {Giannetti},\ and\ \citenamefont {Banfi}}]{Gandolfi2019}%
  \BibitemOpen
  \bibfield  {author} {\bibinfo {author} {\bibfnamefont {Marco}\ \bibnamefont
  {Gandolfi}}, \bibinfo {author} {\bibfnamefont {Giulio}\ \bibnamefont
  {Benetti}}, \bibinfo {author} {\bibfnamefont {Christ}\ \bibnamefont
  {Glorieux}}, \bibinfo {author} {\bibfnamefont {Claudio}\ \bibnamefont
  {Giannetti}}, \ and\ \bibinfo {author} {\bibfnamefont {Francesco}\
  \bibnamefont {Banfi}},\ }\bibfield  {title} {\enquote {\bibinfo {title}
  {Accessing temperature waves: a dispersion relation perspective},}\
  }\href@noop {} {\bibfield  {journal} {\bibinfo  {journal} {International
  Journal of Heat and Mass Transfer}\ }\textbf {\bibinfo {volume} {143}},\
  \bibinfo {pages} {118553} (\bibinfo {year} {2019})}\BibitemShut {NoStop}%
\bibitem [{\citenamefont {Huberman}\ \emph {et~al.}(2019)\citenamefont
  {Huberman}, \citenamefont {Duncan}, \citenamefont {Chen}, \citenamefont
  {Song}, \citenamefont {Chiloyan}, \citenamefont {Ding}, \citenamefont
  {Maznev}, \citenamefont {Chen},\ and\ \citenamefont {Nelson}}]{Huberman2019}%
  \BibitemOpen
  \bibfield  {author} {\bibinfo {author} {\bibfnamefont {S.}~\bibnamefont
  {Huberman}}, \bibinfo {author} {\bibfnamefont {R.A.}\ \bibnamefont {Duncan}},
  \bibinfo {author} {\bibfnamefont {K.}~\bibnamefont {Chen}}, \bibinfo {author}
  {\bibfnamefont {B.}~\bibnamefont {Song}}, \bibinfo {author} {\bibfnamefont
  {V.}~\bibnamefont {Chiloyan}}, \bibinfo {author} {\bibfnamefont
  {Z.}~\bibnamefont {Ding}}, \bibinfo {author} {\bibfnamefont {A.A.}\
  \bibnamefont {Maznev}}, \bibinfo {author} {\bibfnamefont {G.}~\bibnamefont
  {Chen}}, \ and\ \bibinfo {author} {\bibfnamefont {K.A.}\ \bibnamefont
  {Nelson}},\ }\bibfield  {title} {\enquote {\bibinfo {title} {{Observation of
  second sound in graphite at temperatures above 100 K}},}\ }\href@noop {}
  {\bibfield  {journal} {\bibinfo  {journal} {Science}\ }\textbf {\bibinfo
  {volume} {364}},\ \bibinfo {pages} {375} (\bibinfo {year}
  {2019})}\BibitemShut {NoStop}%
\bibitem [{\citenamefont {Zhang}\ \emph {et~al.}(2020)\citenamefont {Zhang},
  \citenamefont {Shi}, \citenamefont {You}, \citenamefont {Tao}, \citenamefont
  {Zhong}, \citenamefont {Cheenicode~Kabeer}, \citenamefont {Maldonado},
  \citenamefont {Oppeneer}, \citenamefont {Bauer}, \citenamefont {Rossnagel},
  \citenamefont {Kapteyn},\ and\ \citenamefont {Murnane}}]{Zhang2020}%
  \BibitemOpen
  \bibfield  {author} {\bibinfo {author} {\bibfnamefont {Yingchao}\
  \bibnamefont {Zhang}}, \bibinfo {author} {\bibfnamefont {Xun}\ \bibnamefont
  {Shi}}, \bibinfo {author} {\bibfnamefont {Wenjing}\ \bibnamefont {You}},
  \bibinfo {author} {\bibfnamefont {Zhensheng}\ \bibnamefont {Tao}}, \bibinfo
  {author} {\bibfnamefont {Yigui}\ \bibnamefont {Zhong}}, \bibinfo {author}
  {\bibfnamefont {Fairoja}\ \bibnamefont {Cheenicode~Kabeer}}, \bibinfo
  {author} {\bibfnamefont {Pablo}\ \bibnamefont {Maldonado}}, \bibinfo {author}
  {\bibfnamefont {Peter~M.}\ \bibnamefont {Oppeneer}}, \bibinfo {author}
  {\bibfnamefont {Michael}\ \bibnamefont {Bauer}}, \bibinfo {author}
  {\bibfnamefont {Kai}\ \bibnamefont {Rossnagel}}, \bibinfo {author}
  {\bibfnamefont {Henry}\ \bibnamefont {Kapteyn}}, \ and\ \bibinfo {author}
  {\bibfnamefont {Margaret}\ \bibnamefont {Murnane}},\ }\bibfield  {title}
  {\enquote {\bibinfo {title} {{Coherent modulation of the electron temperature
  and electron{\textendash}phonon couplings in a 2D material}},}\ }\href
  {\doibase 10.1073/pnas.1917341117} {\bibfield  {journal} {\bibinfo  {journal}
  {Proceedings of the National Academy of Sciences}\ }\textbf {\bibinfo
  {volume} {117}},\ \bibinfo {pages} {8788--8793} (\bibinfo {year}
  {2020})}\BibitemShut {NoStop}%
\bibitem [{\citenamefont {Gandolfi}\ \emph {et~al.}(2020)\citenamefont
  {Gandolfi}, \citenamefont {Giannetti},\ and\ \citenamefont
  {Banfi}}]{Gandolfi2020}%
  \BibitemOpen
  \bibfield  {author} {\bibinfo {author} {\bibfnamefont {Marco}\ \bibnamefont
  {Gandolfi}}, \bibinfo {author} {\bibfnamefont {Claudio}\ \bibnamefont
  {Giannetti}}, \ and\ \bibinfo {author} {\bibfnamefont {Francesco}\
  \bibnamefont {Banfi}},\ }\bibfield  {title} {\enquote {\bibinfo {title}
  {Temperonic crystal: A superlattice for temperature waves in graphene},}\
  }\href {\doibase 10.1103/PhysRevLett.125.265901} {\bibfield  {journal}
  {\bibinfo  {journal} {Phys. Rev. Lett.}\ }\textbf {\bibinfo {volume} {125}},\
  \bibinfo {pages} {265901} (\bibinfo {year} {2020})}\BibitemShut {NoStop}%
\bibitem [{\citenamefont {Cepellotti}\ and\ \citenamefont
  {Marzari}(2017)}]{Cepellotti2017}%
  \BibitemOpen
  \bibfield  {author} {\bibinfo {author} {\bibfnamefont {Andrea}\ \bibnamefont
  {Cepellotti}}\ and\ \bibinfo {author} {\bibfnamefont {Nicola}\ \bibnamefont
  {Marzari}},\ }\bibfield  {title} {\enquote {\bibinfo {title} {Transport waves
  as crystal excitations},}\ }\href {\doibase
  10.1103/PhysRevMaterials.1.045406} {\bibfield  {journal} {\bibinfo  {journal}
  {Phys. Rev. Materials}\ }\textbf {\bibinfo {volume} {1}},\ \bibinfo {pages}
  {045406} (\bibinfo {year} {2017})}\BibitemShut {NoStop}%
\bibitem [{\citenamefont {Torres}\ \emph {et~al.}(2019)\citenamefont {Torres},
  \citenamefont {Alvarez}, \citenamefont {Cartoix{\`{a}}},\ and\ \citenamefont
  {Rurali}}]{Torres2019}%
  \BibitemOpen
  \bibfield  {author} {\bibinfo {author} {\bibfnamefont {Pol}\ \bibnamefont
  {Torres}}, \bibinfo {author} {\bibfnamefont {Francesc~Xavier}\ \bibnamefont
  {Alvarez}}, \bibinfo {author} {\bibfnamefont {Xavier}\ \bibnamefont
  {Cartoix{\`{a}}}}, \ and\ \bibinfo {author} {\bibfnamefont {Riccardo}\
  \bibnamefont {Rurali}},\ }\bibfield  {title} {\enquote {\bibinfo {title}
  {Thermal conductivity and phonon hydrodynamics in transition metal
  dichalcogenides from first-principles},}\ }\href {\doibase
  10.1088/2053-1583/ab0c31} {\bibfield  {journal} {\bibinfo  {journal} {2D
  Materials}\ }\textbf {\bibinfo {volume} {6}},\ \bibinfo {pages} {035002}
  (\bibinfo {year} {2019})}\BibitemShut {NoStop}%
\bibitem [{\citenamefont {Machida}\ \emph {et~al.}(2020)\citenamefont
  {Machida}, \citenamefont {Matsumoto}, \citenamefont {Isono},\ and\
  \citenamefont {Behnia}}]{Machida2020}%
  \BibitemOpen
  \bibfield  {author} {\bibinfo {author} {\bibfnamefont {Yo}~\bibnamefont
  {Machida}}, \bibinfo {author} {\bibfnamefont {Nayuta}\ \bibnamefont
  {Matsumoto}}, \bibinfo {author} {\bibfnamefont {Takayuki}\ \bibnamefont
  {Isono}}, \ and\ \bibinfo {author} {\bibfnamefont {Kamran}\ \bibnamefont
  {Behnia}},\ }\bibfield  {title} {\enquote {\bibinfo {title} {Phonon
  hydrodynamics and ultrahigh{\textendash}room-temperature thermal conductivity
  in thin graphite},}\ }\href {\doibase 10.1126/science.aaz8043} {\bibfield
  {journal} {\bibinfo  {journal} {Science}\ }\textbf {\bibinfo {volume}
  {367}},\ \bibinfo {pages} {309--312} (\bibinfo {year} {2020})}\BibitemShut
  {NoStop}%
\bibitem [{\citenamefont {Simoncelli}\ \emph {et~al.}(2020)\citenamefont
  {Simoncelli}, \citenamefont {Marzari},\ and\ \citenamefont
  {Cepellotti}}]{simoncelli2020}%
  \BibitemOpen
  \bibfield  {author} {\bibinfo {author} {\bibfnamefont {Michele}\ \bibnamefont
  {Simoncelli}}, \bibinfo {author} {\bibfnamefont {Nicola}\ \bibnamefont
  {Marzari}}, \ and\ \bibinfo {author} {\bibfnamefont {Andrea}\ \bibnamefont
  {Cepellotti}},\ }\bibfield  {title} {\enquote {\bibinfo {title}
  {{Generalization of Fourier's Law into Viscous Heat Equations}},}\ }\href
  {\doibase 10.1103/PhysRevX.10.011019} {\bibfield  {journal} {\bibinfo
  {journal} {Phys. Rev. X}\ }\textbf {\bibinfo {volume} {10}},\ \bibinfo
  {pages} {011019} (\bibinfo {year} {2020})}\BibitemShut {NoStop}%
\bibitem [{\citenamefont {Gandolfi}\ \emph {et~al.}(2017)\citenamefont
  {Gandolfi}, \citenamefont {Celardo}, \citenamefont {Borgonovi}, \citenamefont
  {Ferrini}, \citenamefont {Avella}, \citenamefont {Banfi},\ and\ \citenamefont
  {Giannetti}}]{Gandolfi2017}%
  \BibitemOpen
  \bibfield  {author} {\bibinfo {author} {\bibfnamefont {Marco}\ \bibnamefont
  {Gandolfi}}, \bibinfo {author} {\bibfnamefont {Luca~Giuseppe}\ \bibnamefont
  {Celardo}}, \bibinfo {author} {\bibfnamefont {Fausto}\ \bibnamefont
  {Borgonovi}}, \bibinfo {author} {\bibfnamefont {Gabriele}\ \bibnamefont
  {Ferrini}}, \bibinfo {author} {\bibfnamefont {Adolfo}\ \bibnamefont
  {Avella}}, \bibinfo {author} {\bibfnamefont {Francesco}\ \bibnamefont
  {Banfi}}, \ and\ \bibinfo {author} {\bibfnamefont {Claudio}\ \bibnamefont
  {Giannetti}},\ }\bibfield  {title} {\enquote {\bibinfo {title} {Emergent
  ultrafast phenomena in correlated oxides and heterostructures},}\ }\href@noop
  {} {\bibfield  {journal} {\bibinfo  {journal} {Phys. Scripta}\ }\textbf
  {\bibinfo {volume} {92}},\ \bibinfo {pages} {034004} (\bibinfo {year}
  {2017})}\BibitemShut {NoStop}%
\bibitem [{\citenamefont {Hartnoll}(2015)}]{Hartnoll2015}%
  \BibitemOpen
  \bibfield  {author} {\bibinfo {author} {\bibfnamefont {Sean~A}\ \bibnamefont
  {Hartnoll}},\ }\bibfield  {title} {\enquote {\bibinfo {title} {{Theory of
  universal incoherent metallic transport}},}\ }\href@noop {} {\bibfield
  {journal} {\bibinfo  {journal} {Nature Physics}\ }\textbf {\bibinfo {volume}
  {11}},\ \bibinfo {pages} {54--61} (\bibinfo {year} {2015})}\BibitemShut
  {NoStop}%
\bibitem [{\citenamefont {Lee}\ \emph {et~al.}(2017)\citenamefont {Lee},
  \citenamefont {Hippalgaonkar}, \citenamefont {Yang}, \citenamefont {Hong},
  \citenamefont {Ko}, \citenamefont {Suh}, \citenamefont {Liu}, \citenamefont
  {Wang}, \citenamefont {Urban}, \citenamefont {Zhang}, \citenamefont {Dames},
  \citenamefont {Hartnoll}, \citenamefont {Delaire},\ and\ \citenamefont
  {Wu}}]{Lee2017}%
  \BibitemOpen
  \bibfield  {author} {\bibinfo {author} {\bibfnamefont {Sangwook}\
  \bibnamefont {Lee}}, \bibinfo {author} {\bibfnamefont {Kedar}\ \bibnamefont
  {Hippalgaonkar}}, \bibinfo {author} {\bibfnamefont {Fan}\ \bibnamefont
  {Yang}}, \bibinfo {author} {\bibfnamefont {Jiawang}\ \bibnamefont {Hong}},
  \bibinfo {author} {\bibfnamefont {Changhyun}\ \bibnamefont {Ko}}, \bibinfo
  {author} {\bibfnamefont {Joonki}\ \bibnamefont {Suh}}, \bibinfo {author}
  {\bibfnamefont {Kai}\ \bibnamefont {Liu}}, \bibinfo {author} {\bibfnamefont
  {Kevin}\ \bibnamefont {Wang}}, \bibinfo {author} {\bibfnamefont {Jeffrey~J.}\
  \bibnamefont {Urban}}, \bibinfo {author} {\bibfnamefont {Xiang}\ \bibnamefont
  {Zhang}}, \bibinfo {author} {\bibfnamefont {Chris}\ \bibnamefont {Dames}},
  \bibinfo {author} {\bibfnamefont {Sean~A.}\ \bibnamefont {Hartnoll}},
  \bibinfo {author} {\bibfnamefont {Olivier}\ \bibnamefont {Delaire}}, \ and\
  \bibinfo {author} {\bibfnamefont {Junqiao}\ \bibnamefont {Wu}},\ }\bibfield
  {title} {\enquote {\bibinfo {title} {Anomalously low electronic thermal
  conductivity in metallic vanadium dioxide},}\ }\href@noop {} {\bibfield
  {journal} {\bibinfo  {journal} {Science}\ }\textbf {\bibinfo {volume}
  {355}},\ \bibinfo {pages} {371--374} (\bibinfo {year} {2017})}\BibitemShut
  {NoStop}%
\bibitem [{\citenamefont {Tokura}\ \emph {et~al.}(2017)\citenamefont {Tokura},
  \citenamefont {Kawasaki},\ and\ \citenamefont {Nagaosa}}]{Tokura2017}%
  \BibitemOpen
  \bibfield  {author} {\bibinfo {author} {\bibfnamefont {Yoshinori}\
  \bibnamefont {Tokura}}, \bibinfo {author} {\bibfnamefont {Masashi}\
  \bibnamefont {Kawasaki}}, \ and\ \bibinfo {author} {\bibfnamefont {Naoto}\
  \bibnamefont {Nagaosa}},\ }\bibfield  {title} {\enquote {\bibinfo {title}
  {{Emergent functions of quantum materials}},}\ }\href {\doibase
  10.1038/nphys4274} {\bibfield  {journal} {\bibinfo  {journal} {Nature
  Physics}\ }\textbf {\bibinfo {volume} {13}},\ \bibinfo {pages} {1056--1068}
  (\bibinfo {year} {2017})}\BibitemShut {NoStop}%
\bibitem [{\citenamefont {Basov}\ \emph {et~al.}(2017)\citenamefont {Basov},
  \citenamefont {Averitt},\ and\ \citenamefont {Hsieh}}]{Basov_NatMat_2017}%
  \BibitemOpen
  \bibfield  {author} {\bibinfo {author} {\bibfnamefont {D~N}\ \bibnamefont
  {Basov}}, \bibinfo {author} {\bibfnamefont {N~D}\ \bibnamefont {Averitt}}, \
  and\ \bibinfo {author} {\bibfnamefont {D}~\bibnamefont {Hsieh}},\ }\bibfield
  {title} {\enquote {\bibinfo {title} {Towards properties on demand in quantum
  materials},}\ }\href@noop {} {\bibfield  {journal} {\bibinfo  {journal} {Nat.
  Mat.}\ }\textbf {\bibinfo {volume} {16}},\ \bibinfo {pages} {1077--1088}
  (\bibinfo {year} {2017})}\BibitemShut {NoStop}%
\bibitem [{\citenamefont {Ordonez-Miranda}\ \emph {et~al.}(2018)\citenamefont
  {Ordonez-Miranda}, \citenamefont {Ezzahri}, \citenamefont {Joulain},
  \citenamefont {Drevillon},\ and\ \citenamefont
  {Alvarado-Gil}}]{Miranda_PRB2018}%
  \BibitemOpen
  \bibfield  {author} {\bibinfo {author} {\bibfnamefont {Jose}\ \bibnamefont
  {Ordonez-Miranda}}, \bibinfo {author} {\bibfnamefont {Youn\`es}\ \bibnamefont
  {Ezzahri}}, \bibinfo {author} {\bibfnamefont {Karl}\ \bibnamefont {Joulain}},
  \bibinfo {author} {\bibfnamefont {J\'er\'emie}\ \bibnamefont {Drevillon}}, \
  and\ \bibinfo {author} {\bibfnamefont {J.~J.}\ \bibnamefont {Alvarado-Gil}},\
  }\bibfield  {title} {\enquote {\bibinfo {title} {{Modeling of the electrical
  conductivity, thermal conductivity, and specific heat capacity of
  ${\mathrm{VO}}_{2}$}},}\ }\href {\doibase 10.1103/PhysRevB.98.075144}
  {\bibfield  {journal} {\bibinfo  {journal} {Phys. Rev. B}\ }\textbf {\bibinfo
  {volume} {98}},\ \bibinfo {pages} {075144} (\bibinfo {year}
  {2018})}\BibitemShut {NoStop}%
\bibitem [{\citenamefont {Cesarini}\ \emph {et~al.}(2019)\citenamefont
  {Cesarini}, \citenamefont {Leahu}, \citenamefont {Belardini}, \citenamefont
  {Centini}, \citenamefont {Voti},\ and\ \citenamefont
  {Sibilia}}]{cesarini2019}%
  \BibitemOpen
  \bibfield  {author} {\bibinfo {author} {\bibfnamefont {Gianmario}\
  \bibnamefont {Cesarini}}, \bibinfo {author} {\bibfnamefont {Grigore}\
  \bibnamefont {Leahu}}, \bibinfo {author} {\bibfnamefont {Alessandro}\
  \bibnamefont {Belardini}}, \bibinfo {author} {\bibfnamefont {Marco}\
  \bibnamefont {Centini}}, \bibinfo {author} {\bibfnamefont {Roberto~Li}\
  \bibnamefont {Voti}}, \ and\ \bibinfo {author} {\bibfnamefont {Concita}\
  \bibnamefont {Sibilia}},\ }\bibfield  {title} {\enquote {\bibinfo {title}
  {Quantitative evaluation of emission properties and thermal hysteresis in the
  mid-infrared for a single thin film of vanadium dioxide on a silicon
  substrate},}\ }\href@noop {} {\bibfield  {journal} {\bibinfo  {journal}
  {International Journal of Thermal Sciences}\ }\textbf {\bibinfo {volume}
  {146}},\ \bibinfo {pages} {106061} (\bibinfo {year} {2019})}\BibitemShut
  {NoStop}%
\bibitem [{\citenamefont {Ben-Abdallah}\ and\ \citenamefont
  {Biehs}(2014)}]{Ben-Abdallah_2014}%
  \BibitemOpen
  \bibfield  {author} {\bibinfo {author} {\bibfnamefont {Philippe}\
  \bibnamefont {Ben-Abdallah}}\ and\ \bibinfo {author} {\bibfnamefont
  {Svend-Age}\ \bibnamefont {Biehs}},\ }\bibfield  {title} {\enquote {\bibinfo
  {title} {Near-field thermal transistor},}\ }\href {\doibase
  10.1103/PhysRevLett.112.044301} {\bibfield  {journal} {\bibinfo  {journal}
  {Phys. Rev. Lett.}\ }\textbf {\bibinfo {volume} {112}},\ \bibinfo {pages}
  {044301} (\bibinfo {year} {2014})}\BibitemShut {NoStop}%
\bibitem [{\citenamefont {Ordonez-Miranda}\ \emph {et~al.}(2019)\citenamefont
  {Ordonez-Miranda}, \citenamefont {Ezzahri}, \citenamefont {Tiburcio-Moreno},
  \citenamefont {Joulain},\ and\ \citenamefont {Drevillon}}]{Miranda_PRL2019}%
  \BibitemOpen
  \bibfield  {author} {\bibinfo {author} {\bibfnamefont {Jose}\ \bibnamefont
  {Ordonez-Miranda}}, \bibinfo {author} {\bibfnamefont {Youn\`es}\ \bibnamefont
  {Ezzahri}}, \bibinfo {author} {\bibfnamefont {Jose~A.}\ \bibnamefont
  {Tiburcio-Moreno}}, \bibinfo {author} {\bibfnamefont {Karl}\ \bibnamefont
  {Joulain}}, \ and\ \bibinfo {author} {\bibfnamefont {J\'er\'emie}\
  \bibnamefont {Drevillon}},\ }\bibfield  {title} {\enquote {\bibinfo {title}
  {Radiative thermal memristor},}\ }\href {\doibase
  10.1103/PhysRevLett.123.025901} {\bibfield  {journal} {\bibinfo  {journal}
  {Phys. Rev. Lett.}\ }\textbf {\bibinfo {volume} {123}},\ \bibinfo {pages}
  {025901} (\bibinfo {year} {2019})}\BibitemShut {NoStop}%
\bibitem [{\citenamefont {Moyer}\ \emph {et~al.}(2013)\citenamefont {Moyer},
  \citenamefont {Eaton},\ and\ \citenamefont {Engel-Herbert}}]{Moyer2013}%
  \BibitemOpen
  \bibfield  {author} {\bibinfo {author} {\bibfnamefont {Jarrett~A.}\
  \bibnamefont {Moyer}}, \bibinfo {author} {\bibfnamefont {Craig}\ \bibnamefont
  {Eaton}}, \ and\ \bibinfo {author} {\bibfnamefont {Roman}\ \bibnamefont
  {Engel-Herbert}},\ }\bibfield  {title} {\enquote {\bibinfo {title} {Highly
  conductive srvo3 as a bottom electrode for functional perovskite oxides},}\
  }\href@noop {} {\bibfield  {journal} {\bibinfo  {journal} {Advanced
  Materials}\ }\textbf {\bibinfo {volume} {25}},\ \bibinfo {pages} {3578--3582}
  (\bibinfo {year} {2013})}\BibitemShut {NoStop}%
\bibitem [{\citenamefont {Zhong}\ \emph {et~al.}(2015)\citenamefont {Zhong},
  \citenamefont {Wallerberger}, \citenamefont {Tomczak}, \citenamefont
  {Taranto}, \citenamefont {Parragh}, \citenamefont {Toschi}, \citenamefont
  {Sangiovanni},\ and\ \citenamefont {Held}}]{Zhong2015}%
  \BibitemOpen
  \bibfield  {author} {\bibinfo {author} {\bibfnamefont {Zhicheng}\
  \bibnamefont {Zhong}}, \bibinfo {author} {\bibfnamefont {Markus}\
  \bibnamefont {Wallerberger}}, \bibinfo {author} {\bibfnamefont {Jan~M.}\
  \bibnamefont {Tomczak}}, \bibinfo {author} {\bibfnamefont {Ciro}\
  \bibnamefont {Taranto}}, \bibinfo {author} {\bibfnamefont {Nicolaus}\
  \bibnamefont {Parragh}}, \bibinfo {author} {\bibfnamefont {Alessandro}\
  \bibnamefont {Toschi}}, \bibinfo {author} {\bibfnamefont {Giorgio}\
  \bibnamefont {Sangiovanni}}, \ and\ \bibinfo {author} {\bibfnamefont
  {Karsten}\ \bibnamefont {Held}},\ }\bibfield  {title} {\enquote {\bibinfo
  {title} {{Electronics with Correlated Oxides:
  ${\mathrm{SrVO}}_{3}/{\mathrm{SrTiO}}_{3}$ as a Mott Transistor}},}\
  }\href@noop {} {\bibfield  {journal} {\bibinfo  {journal} {Phys. Rev. Lett.}\
  }\textbf {\bibinfo {volume} {114}},\ \bibinfo {pages} {246401} (\bibinfo
  {year} {2015})}\BibitemShut {NoStop}%
\bibitem [{\citenamefont {Zhang}\ \emph {et~al.}(2015)\citenamefont {Zhang},
  \citenamefont {Zhou}, \citenamefont {Guo}, \citenamefont {Zhao},
  \citenamefont {Barnes}, \citenamefont {Zhang}, \citenamefont {Eaton},
  \citenamefont {Zheng}, \citenamefont {Brahlek}, \citenamefont {Haneef},
  \citenamefont {Podraza}, \citenamefont {Chan}, \citenamefont {Gopalan},
  \citenamefont {Rabe},\ and\ \citenamefont {Engel-Herbert}}]{Zhang2015}%
  \BibitemOpen
  \bibfield  {author} {\bibinfo {author} {\bibfnamefont {Lei}\ \bibnamefont
  {Zhang}}, \bibinfo {author} {\bibfnamefont {Yuanjun}\ \bibnamefont {Zhou}},
  \bibinfo {author} {\bibfnamefont {Lu}~\bibnamefont {Guo}}, \bibinfo {author}
  {\bibfnamefont {Weiwei}\ \bibnamefont {Zhao}}, \bibinfo {author}
  {\bibfnamefont {Anna}\ \bibnamefont {Barnes}}, \bibinfo {author}
  {\bibfnamefont {Hai-Tian}\ \bibnamefont {Zhang}}, \bibinfo {author}
  {\bibfnamefont {Craig}\ \bibnamefont {Eaton}}, \bibinfo {author}
  {\bibfnamefont {Yuanxia}\ \bibnamefont {Zheng}}, \bibinfo {author}
  {\bibfnamefont {Matthew}\ \bibnamefont {Brahlek}}, \bibinfo {author}
  {\bibfnamefont {Hamna~F.}\ \bibnamefont {Haneef}}, \bibinfo {author}
  {\bibfnamefont {Nikolas~J.}\ \bibnamefont {Podraza}}, \bibinfo {author}
  {\bibfnamefont {Moses H.~W.}\ \bibnamefont {Chan}}, \bibinfo {author}
  {\bibfnamefont {Venkatraman}\ \bibnamefont {Gopalan}}, \bibinfo {author}
  {\bibfnamefont {Karin~M.}\ \bibnamefont {Rabe}}, \ and\ \bibinfo {author}
  {\bibfnamefont {Roman}\ \bibnamefont {Engel-Herbert}},\ }\bibfield  {title}
  {\enquote {\bibinfo {title} {Correlated metals as transparent conductors},}\
  }\href@noop {} {\bibfield  {journal} {\bibinfo  {journal} {Nature Materials}\
  }\textbf {\bibinfo {volume} {15}},\ \bibinfo {pages} {204--210} (\bibinfo
  {year} {2015})}\BibitemShut {NoStop}%
\bibitem [{\citenamefont {Fabrizio}(2012)}]{michele_review}%
  \BibitemOpen
  \bibfield  {author} {\bibinfo {author} {\bibfnamefont {Michele}\ \bibnamefont
  {Fabrizio}},\ }\bibfield  {title} {\enquote {\bibinfo {title} {The
  out-of-equilibrium time-dependent gutzwiller approximation},}\ }\href
  {\doibase 10.1007/978-94-007-4984-916} {\bibfield  {journal} {\bibinfo
  {journal} {NATO Science for Peace and Security Series B: Physics and
  Biophysics}\ } (\bibinfo {year} {2012}),\
  10.1007/978-94-007-4984-916}\BibitemShut {NoStop}%
\bibitem [{\citenamefont {Mazza}\ \emph {et~al.}(2015)\citenamefont {Mazza},
  \citenamefont {Amaricci}, \citenamefont {Capone},\ and\ \citenamefont
  {Fabrizio}}]{giacomo_phd_prb}%
  \BibitemOpen
  \bibfield  {author} {\bibinfo {author} {\bibfnamefont {G.}~\bibnamefont
  {Mazza}}, \bibinfo {author} {\bibfnamefont {A.}~\bibnamefont {Amaricci}},
  \bibinfo {author} {\bibfnamefont {M.}~\bibnamefont {Capone}}, \ and\ \bibinfo
  {author} {\bibfnamefont {M.}~\bibnamefont {Fabrizio}},\ }\bibfield  {title}
  {\enquote {\bibinfo {title} {Electronic transport and dynamics in correlated
  heterostructures},}\ }\href {\doibase 10.1103/PhysRevB.91.195124} {\bibfield
  {journal} {\bibinfo  {journal} {Phys. Rev. B}\ }\textbf {\bibinfo {volume}
  {91}},\ \bibinfo {pages} {195124} (\bibinfo {year} {2015})}\BibitemShut
  {NoStop}%
\bibitem [{\citenamefont {Georges}\ \emph {et~al.}(1996)\citenamefont
  {Georges}, \citenamefont {Kotliar}, \citenamefont {Krauth},\ and\
  \citenamefont {Rozenberg}}]{neq_dmft_review}%
  \BibitemOpen
  \bibfield  {author} {\bibinfo {author} {\bibfnamefont {Antoine}\ \bibnamefont
  {Georges}}, \bibinfo {author} {\bibfnamefont {Gabriel}\ \bibnamefont
  {Kotliar}}, \bibinfo {author} {\bibfnamefont {Werner}\ \bibnamefont
  {Krauth}}, \ and\ \bibinfo {author} {\bibfnamefont {Marcelo~J.}\ \bibnamefont
  {Rozenberg}},\ }\bibfield  {title} {\enquote {\bibinfo {title} {Dynamical
  mean-field theory of strongly correlated fermion systems and the limit of
  infinite dimensions},}\ }\href@noop {} {\bibfield  {journal} {\bibinfo
  {journal} {Rev. Mod. Phys.}\ }\textbf {\bibinfo {volume} {68}},\ \bibinfo
  {pages} {13--125} (\bibinfo {year} {1996})}\BibitemShut {NoStop}%
\bibitem [{Note1()}]{Note1}%
  \BibitemOpen
  \bibinfo {note} {This result is confirmed using different fitting procedures
  in which $T_h$ is either considered a fixed parameter or fitting
  parameter.}\BibitemShut {Stop}%
\bibitem [{\citenamefont {Mingo}\ and\ \citenamefont
  {Broido}(2005)}]{Mingo2005}%
  \BibitemOpen
  \bibfield  {author} {\bibinfo {author} {\bibfnamefont {N.}~\bibnamefont
  {Mingo}}\ and\ \bibinfo {author} {\bibfnamefont {D.~A.}\ \bibnamefont
  {Broido}},\ }\bibfield  {title} {\enquote {\bibinfo {title} {Carbon nanotube
  ballistic thermal conductance and its limits},}\ }\href@noop {} {\bibfield
  {journal} {\bibinfo  {journal} {Phys. Rev. Lett.}\ }\textbf {\bibinfo
  {volume} {95}},\ \bibinfo {pages} {096105} (\bibinfo {year}
  {2005})}\BibitemShut {NoStop}%
\bibitem [{\citenamefont {Mu{\~{n}}oz}\ \emph {et~al.}(2010)\citenamefont
  {Mu{\~{n}}oz}, \citenamefont {Lu},\ and\ \citenamefont
  {Yakobson}}]{Munoz2010}%
  \BibitemOpen
  \bibfield  {author} {\bibinfo {author} {\bibfnamefont {Enrique}\ \bibnamefont
  {Mu{\~{n}}oz}}, \bibinfo {author} {\bibfnamefont {Jianxin}\ \bibnamefont
  {Lu}}, \ and\ \bibinfo {author} {\bibfnamefont {Boris~I}\ \bibnamefont
  {Yakobson}},\ }\bibfield  {title} {\enquote {\bibinfo {title} {{Ballistic
  Thermal Conductance of Graphene Ribbons}},}\ }\href@noop {} {\bibfield
  {journal} {\bibinfo  {journal} {Nano Letters}\ }\textbf {\bibinfo {volume}
  {10}},\ \bibinfo {pages} {1652--1656} (\bibinfo {year} {2010})}\BibitemShut
  {NoStop}%
\bibitem [{\citenamefont {Bae}\ \emph {et~al.}(2013)\citenamefont {Bae},
  \citenamefont {Li}, \citenamefont {Aksamija}, \citenamefont {Martin},
  \citenamefont {Xiong}, \citenamefont {Ong}, \citenamefont {Knezevic},\ and\
  \citenamefont {Pop}}]{Bae2013}%
  \BibitemOpen
  \bibfield  {author} {\bibinfo {author} {\bibfnamefont {Myung-Ho}\
  \bibnamefont {Bae}}, \bibinfo {author} {\bibfnamefont {Zuanyi}\ \bibnamefont
  {Li}}, \bibinfo {author} {\bibfnamefont {Zlatan}\ \bibnamefont {Aksamija}},
  \bibinfo {author} {\bibfnamefont {Pierre~N}\ \bibnamefont {Martin}}, \bibinfo
  {author} {\bibfnamefont {Feng}\ \bibnamefont {Xiong}}, \bibinfo {author}
  {\bibfnamefont {Zhun-Yong}\ \bibnamefont {Ong}}, \bibinfo {author}
  {\bibfnamefont {Irena}\ \bibnamefont {Knezevic}}, \ and\ \bibinfo {author}
  {\bibfnamefont {Eric}\ \bibnamefont {Pop}},\ }\bibfield  {title} {\enquote
  {\bibinfo {title} {{Ballistic to diffusive crossover of heat flow in graphene
  ribbons}},}\ }\href {\doibase 10.1038/ncomms2755} {\bibfield  {journal}
  {\bibinfo  {journal} {Nature Communications}\ }\textbf {\bibinfo {volume}
  {4}},\ \bibinfo {pages} {1734} (\bibinfo {year} {2013})}\BibitemShut
  {NoStop}%
\bibitem [{\citenamefont {Caddeo}\ \emph {et~al.}(2017)\citenamefont {Caddeo},
  \citenamefont {Melis}, \citenamefont {Ronchi}, \citenamefont {Giannetti},
  \citenamefont {Ferrini}, \citenamefont {Rurali}, \citenamefont {Colombo},\
  and\ \citenamefont {Banfi}}]{Caddeo2017}%
  \BibitemOpen
  \bibfield  {author} {\bibinfo {author} {\bibfnamefont {Claudia}\ \bibnamefont
  {Caddeo}}, \bibinfo {author} {\bibfnamefont {Claudio}\ \bibnamefont {Melis}},
  \bibinfo {author} {\bibfnamefont {Andrea}\ \bibnamefont {Ronchi}}, \bibinfo
  {author} {\bibfnamefont {Claudio}\ \bibnamefont {Giannetti}}, \bibinfo
  {author} {\bibfnamefont {Gabriele}\ \bibnamefont {Ferrini}}, \bibinfo
  {author} {\bibfnamefont {Riccardo}\ \bibnamefont {Rurali}}, \bibinfo {author}
  {\bibfnamefont {Luciano}\ \bibnamefont {Colombo}}, \ and\ \bibinfo {author}
  {\bibfnamefont {Francesco}\ \bibnamefont {Banfi}},\ }\bibfield  {title}
  {\enquote {\bibinfo {title} {Thermal boundary resistance from transient
  nanocalorimetry: A multiscale modeling approach},}\ }\href@noop {} {\bibfield
   {journal} {\bibinfo  {journal} {Phys. Rev. B}\ }\textbf {\bibinfo {volume}
  {95}},\ \bibinfo {pages} {085306} (\bibinfo {year} {2017})}\BibitemShut
  {NoStop}%
\bibitem [{\citenamefont {Tzou}(2014)}]{Tzou2014}%
  \BibitemOpen
  \bibfield  {author} {\bibinfo {author} {\bibfnamefont {Da~Yu}\ \bibnamefont
  {Tzou}},\ }\href@noop {} {\emph {\bibinfo {title} {Macro-to microscale heat
  transfer: the lagging behavior}}}\ (\bibinfo  {publisher} {John Wiley \&
  Sons},\ \bibinfo {year} {2014})\BibitemShut {NoStop}%
\bibitem [{\citenamefont {Aizaki}\ \emph {et~al.}(2012)\citenamefont {Aizaki},
  \citenamefont {Yoshida}, \citenamefont {Yoshimatsu}, \citenamefont
  {Takizawa}, \citenamefont {Minohara}, \citenamefont {Ideta}, \citenamefont
  {Fujimori}, \citenamefont {Gupta}, \citenamefont {Mahadevan}, \citenamefont
  {Horiba}, \citenamefont {Kumigashira},\ and\ \citenamefont
  {Oshima}}]{Aizaki2012}%
  \BibitemOpen
  \bibfield  {author} {\bibinfo {author} {\bibfnamefont {S.}~\bibnamefont
  {Aizaki}}, \bibinfo {author} {\bibfnamefont {T.}~\bibnamefont {Yoshida}},
  \bibinfo {author} {\bibfnamefont {K.}~\bibnamefont {Yoshimatsu}}, \bibinfo
  {author} {\bibfnamefont {M.}~\bibnamefont {Takizawa}}, \bibinfo {author}
  {\bibfnamefont {M.}~\bibnamefont {Minohara}}, \bibinfo {author}
  {\bibfnamefont {S.}~\bibnamefont {Ideta}}, \bibinfo {author} {\bibfnamefont
  {A.}~\bibnamefont {Fujimori}}, \bibinfo {author} {\bibfnamefont
  {K.}~\bibnamefont {Gupta}}, \bibinfo {author} {\bibfnamefont
  {P.}~\bibnamefont {Mahadevan}}, \bibinfo {author} {\bibfnamefont
  {K.}~\bibnamefont {Horiba}}, \bibinfo {author} {\bibfnamefont
  {H.}~\bibnamefont {Kumigashira}}, \ and\ \bibinfo {author} {\bibfnamefont
  {M.}~\bibnamefont {Oshima}},\ }\bibfield  {title} {\enquote {\bibinfo {title}
  {{Self-Energy on the Low- to High-Energy Electronic Structure of Correlated
  Metal ${\mathrm{SrVO}}_{3}$}},}\ }\href@noop {} {\bibfield  {journal}
  {\bibinfo  {journal} {Phys. Rev. Lett.}\ }\textbf {\bibinfo {volume} {109}},\
  \bibinfo {pages} {056401} (\bibinfo {year} {2012})}\BibitemShut {NoStop}%
\bibitem [{\citenamefont {Inoue}\ \emph {et~al.}(1998)\citenamefont {Inoue},
  \citenamefont {Goto}, \citenamefont {Makino}, \citenamefont {Hussey},\ and\
  \citenamefont {Ishikawa}}]{Inoue1998}%
  \BibitemOpen
  \bibfield  {author} {\bibinfo {author} {\bibfnamefont {I.~H.}\ \bibnamefont
  {Inoue}}, \bibinfo {author} {\bibfnamefont {O.}~\bibnamefont {Goto}},
  \bibinfo {author} {\bibfnamefont {H.}~\bibnamefont {Makino}}, \bibinfo
  {author} {\bibfnamefont {N.~E.}\ \bibnamefont {Hussey}}, \ and\ \bibinfo
  {author} {\bibfnamefont {M.}~\bibnamefont {Ishikawa}},\ }\bibfield  {title}
  {\enquote {\bibinfo {title} {{Bandwidth control in a perovskite-type
  ${3d}^{1}$-correlated metal
  ${\mathrm{Ca}}_{1\ensuremath{-}x}{\mathrm{Sr}}_{x}{\mathrm{VO}}_{3}.$ I.
  Evolution of the electronic properties and effective mass}},}\ }\href@noop {}
  {\bibfield  {journal} {\bibinfo  {journal} {Phys. Rev. B}\ }\textbf {\bibinfo
  {volume} {58}},\ \bibinfo {pages} {4372--4383} (\bibinfo {year}
  {1998})}\BibitemShut {NoStop}%
\bibitem [{\citenamefont {Karrasch}\ \emph {et~al.}(2016)\citenamefont
  {Karrasch}, \citenamefont {Kennes},\ and\ \citenamefont
  {Heidrich-Meisner}}]{Karrasch2016}%
  \BibitemOpen
  \bibfield  {author} {\bibinfo {author} {\bibfnamefont {C.}~\bibnamefont
  {Karrasch}}, \bibinfo {author} {\bibfnamefont {D.~M.}\ \bibnamefont
  {Kennes}}, \ and\ \bibinfo {author} {\bibfnamefont {F.}~\bibnamefont
  {Heidrich-Meisner}},\ }\bibfield  {title} {\enquote {\bibinfo {title}
  {{Thermal Conductivity of the One-Dimensional Fermi-Hubbard Model}},}\ }\href
  {\doibase 10.1103/PhysRevLett.117.116401} {\bibfield  {journal} {\bibinfo
  {journal} {Phys. Rev. Lett.}\ }\textbf {\bibinfo {volume} {117}},\ \bibinfo
  {pages} {116401} (\bibinfo {year} {2016})}\BibitemShut {NoStop}%
\bibitem [{\citenamefont {Giannetti}\ \emph {et~al.}(2016)\citenamefont
  {Giannetti}, \citenamefont {Capone}, \citenamefont {Fausti}, \citenamefont
  {Fabrizio}, \citenamefont {Parmigiani},\ and\ \citenamefont
  {Mihailovic}}]{Giannetti2016}%
  \BibitemOpen
  \bibfield  {author} {\bibinfo {author} {\bibfnamefont {C.}~\bibnamefont
  {Giannetti}}, \bibinfo {author} {\bibfnamefont {M.}~\bibnamefont {Capone}},
  \bibinfo {author} {\bibfnamefont {D.}~\bibnamefont {Fausti}}, \bibinfo
  {author} {\bibfnamefont {M.}~\bibnamefont {Fabrizio}}, \bibinfo {author}
  {\bibfnamefont {F.}~\bibnamefont {Parmigiani}}, \ and\ \bibinfo {author}
  {\bibfnamefont {D.}~\bibnamefont {Mihailovic}},\ }\bibfield  {title}
  {\enquote {\bibinfo {title} {Ultrafast optical spectroscopy of strongly
  correlated materials and high-temperature superconductors: a non-equilibrium
  approach},}\ }\href@noop {} {\bibfield  {journal} {\bibinfo  {journal} {Adv.
  Phys.}\ }\textbf {\bibinfo {volume} {65}},\ \bibinfo {pages} {58} (\bibinfo
  {year} {2016})}\BibitemShut {NoStop}%
\bibitem [{\citenamefont {Hwang}\ \emph {et~al.}(2012)\citenamefont {Hwang},
  \citenamefont {Iwasa}, \citenamefont {Kawasaki}, \citenamefont {Keimer},
  \citenamefont {Nagaosa},\ and\ \citenamefont {Tokura}}]{Hwang2012}%
  \BibitemOpen
  \bibfield  {author} {\bibinfo {author} {\bibfnamefont {H.~Y.}\ \bibnamefont
  {Hwang}}, \bibinfo {author} {\bibfnamefont {Y.}~\bibnamefont {Iwasa}},
  \bibinfo {author} {\bibfnamefont {M.}~\bibnamefont {Kawasaki}}, \bibinfo
  {author} {\bibfnamefont {B.}~\bibnamefont {Keimer}}, \bibinfo {author}
  {\bibfnamefont {N.}~\bibnamefont {Nagaosa}}, \ and\ \bibinfo {author}
  {\bibfnamefont {Y.}~\bibnamefont {Tokura}},\ }\bibfield  {title} {\enquote
  {\bibinfo {title} {Emergent phenomena at oxide interfaces},}\ }\href@noop {}
  {\bibfield  {journal} {\bibinfo  {journal} {Nature Materials}\ }\textbf
  {\bibinfo {volume} {11}},\ \bibinfo {pages} {103--113} (\bibinfo {year}
  {2012})}\BibitemShut {NoStop}%
\bibitem [{\citenamefont {Cao}\ \emph {et~al.}(2018)\citenamefont {Cao},
  \citenamefont {Fatemi}, \citenamefont {Demir}, \citenamefont {Fang},
  \citenamefont {Tomarken}, \citenamefont {Luo}, \citenamefont
  {Sanchez-Yamagishi}, \citenamefont {Watanabe}, \citenamefont {Taniguchi},
  \citenamefont {Kaxiras},\ and\ \citenamefont {et~al.}}]{Cao2018}%
  \BibitemOpen
  \bibfield  {author} {\bibinfo {author} {\bibfnamefont {Yuan}\ \bibnamefont
  {Cao}}, \bibinfo {author} {\bibfnamefont {Valla}\ \bibnamefont {Fatemi}},
  \bibinfo {author} {\bibfnamefont {Ahmet}\ \bibnamefont {Demir}}, \bibinfo
  {author} {\bibfnamefont {Shiang}\ \bibnamefont {Fang}}, \bibinfo {author}
  {\bibfnamefont {Spencer~L.}\ \bibnamefont {Tomarken}}, \bibinfo {author}
  {\bibfnamefont {Jason~Y.}\ \bibnamefont {Luo}}, \bibinfo {author}
  {\bibfnamefont {Javier~D.}\ \bibnamefont {Sanchez-Yamagishi}}, \bibinfo
  {author} {\bibfnamefont {Kenji}\ \bibnamefont {Watanabe}}, \bibinfo {author}
  {\bibfnamefont {Takashi}\ \bibnamefont {Taniguchi}}, \bibinfo {author}
  {\bibfnamefont {Efthimios}\ \bibnamefont {Kaxiras}}, \ and\ \bibinfo {author}
  {\bibnamefont {et~al.}},\ }\bibfield  {title} {\enquote {\bibinfo {title}
  {Correlated insulator behaviour at half-filling in magic-angle graphene
  superlattices},}\ }\href {\doibase 10.1038/nature26154} {\bibfield  {journal}
  {\bibinfo  {journal} {Nature}\ }\textbf {\bibinfo {volume} {556}},\ \bibinfo
  {pages} {80?84} (\bibinfo {year} {2018})}\BibitemShut {NoStop}%
\bibitem [{\citenamefont {Seyler}\ \emph {et~al.}(2019)\citenamefont {Seyler},
  \citenamefont {Rivera}, \citenamefont {Yu}, \citenamefont {Wilson},
  \citenamefont {Ray}, \citenamefont {Mandrus}, \citenamefont {Yan},
  \citenamefont {Yao},\ and\ \citenamefont {Xu}}]{seyler2019}%
  \BibitemOpen
  \bibfield  {author} {\bibinfo {author} {\bibfnamefont {Kyle~L}\ \bibnamefont
  {Seyler}}, \bibinfo {author} {\bibfnamefont {Pasqual}\ \bibnamefont
  {Rivera}}, \bibinfo {author} {\bibfnamefont {Hongyi}\ \bibnamefont {Yu}},
  \bibinfo {author} {\bibfnamefont {Nathan~P}\ \bibnamefont {Wilson}}, \bibinfo
  {author} {\bibfnamefont {Essance~L}\ \bibnamefont {Ray}}, \bibinfo {author}
  {\bibfnamefont {David~G}\ \bibnamefont {Mandrus}}, \bibinfo {author}
  {\bibfnamefont {Jiaqiang}\ \bibnamefont {Yan}}, \bibinfo {author}
  {\bibfnamefont {Wang}\ \bibnamefont {Yao}}, \ and\ \bibinfo {author}
  {\bibfnamefont {Xiaodong}\ \bibnamefont {Xu}},\ }\bibfield  {title} {\enquote
  {\bibinfo {title} {{Signatures of moir{\'e}-trapped valley excitons in
  MoSe$_2$/WSe$_2$ heterobilayers}},}\ }\href@noop {} {\bibfield  {journal}
  {\bibinfo  {journal} {Nature}\ }\textbf {\bibinfo {volume} {567}},\ \bibinfo
  {pages} {66--70} (\bibinfo {year} {2019})}\BibitemShut {NoStop}%
\bibitem [{\citenamefont {Lu}\ \emph {et~al.}(2019)\citenamefont {Lu},
  \citenamefont {Stepanov}, \citenamefont {Yang}, \citenamefont {Xie},
  \citenamefont {Aamir}, \citenamefont {Das}, \citenamefont {Urgell},
  \citenamefont {Watanabe}, \citenamefont {Taniguchi}, \citenamefont {Zhang}
  \emph {et~al.}}]{lu2019}%
  \BibitemOpen
  \bibfield  {author} {\bibinfo {author} {\bibfnamefont {Xiaobo}\ \bibnamefont
  {Lu}}, \bibinfo {author} {\bibfnamefont {Petr}\ \bibnamefont {Stepanov}},
  \bibinfo {author} {\bibfnamefont {Wei}\ \bibnamefont {Yang}}, \bibinfo
  {author} {\bibfnamefont {Ming}\ \bibnamefont {Xie}}, \bibinfo {author}
  {\bibfnamefont {Mohammed~Ali}\ \bibnamefont {Aamir}}, \bibinfo {author}
  {\bibfnamefont {Ipsita}\ \bibnamefont {Das}}, \bibinfo {author}
  {\bibfnamefont {Carles}\ \bibnamefont {Urgell}}, \bibinfo {author}
  {\bibfnamefont {Kenji}\ \bibnamefont {Watanabe}}, \bibinfo {author}
  {\bibfnamefont {Takashi}\ \bibnamefont {Taniguchi}}, \bibinfo {author}
  {\bibfnamefont {Guangyu}\ \bibnamefont {Zhang}},  \emph {et~al.},\ }\bibfield
   {title} {\enquote {\bibinfo {title} {Superconductors, orbital magnets and
  correlated states in magic-angle bilayer graphene},}\ }\href@noop {}
  {\bibfield  {journal} {\bibinfo  {journal} {Nature}\ }\textbf {\bibinfo
  {volume} {574}},\ \bibinfo {pages} {653--657} (\bibinfo {year}
  {2019})}\BibitemShut {NoStop}%
\end{thebibliography}%

\begin{widetext}
\setcounter{equation}{0}
\renewcommand{\theequation}{S\arabic{equation}}

\setcounter{figure}{0}
\renewcommand{\thefigure}{S\arabic{figure}}
\renewcommand{\thetable}{S\arabic{table}}

\clearpage
\newpage
\section*{Supplementary Information for Thermal dynamics and electronic temperature waves in layered correlated materials}

\subsection*{Supplementary figures}
\begin{figure}[h]
\includegraphics[width=0.45\textwidth]{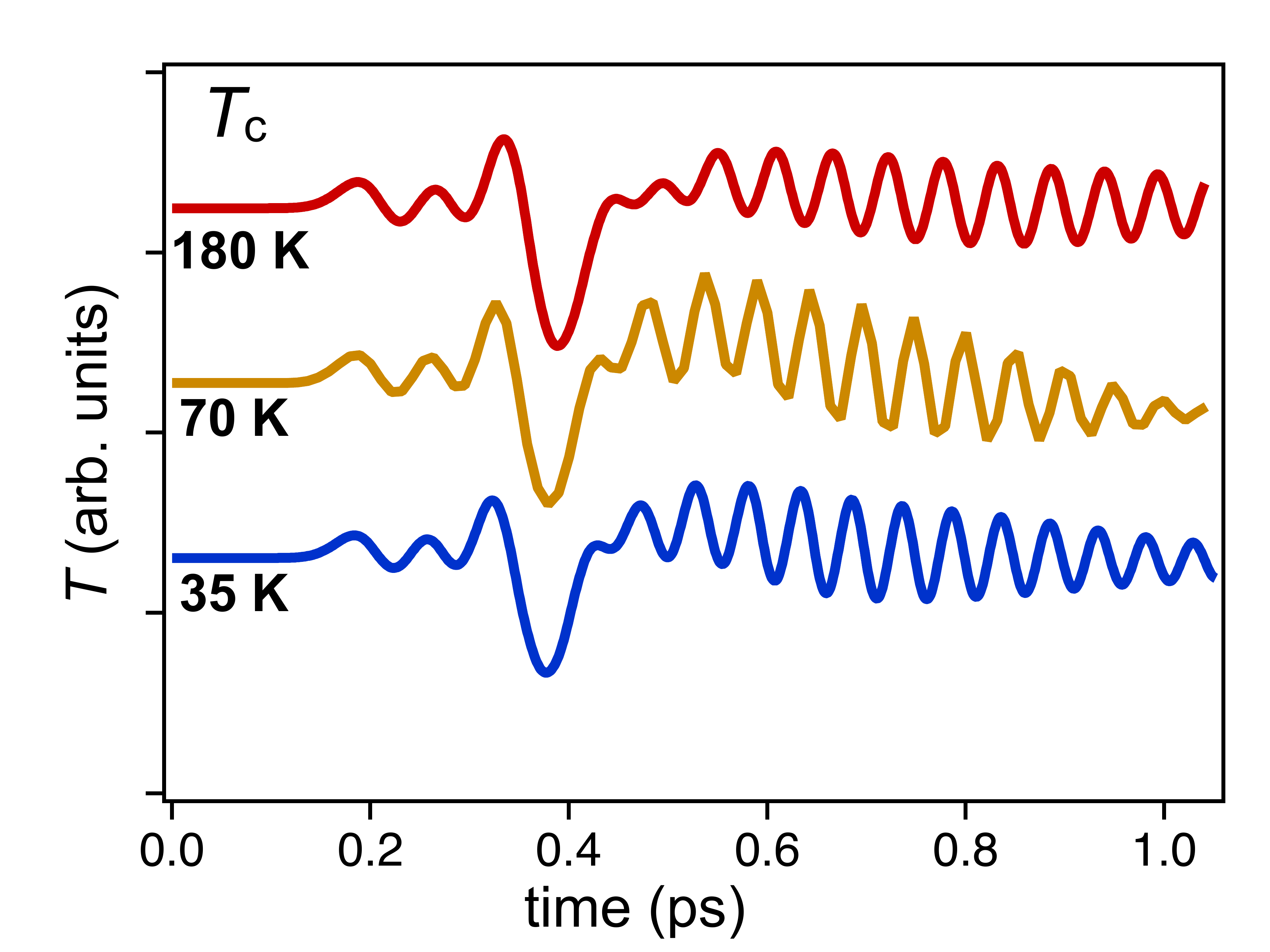}
\caption{Temperature $T(t)$ of the quasi-equilibrium electronic distribution calculated at the 15$^{th}$ layer for different values of the initial base temperature $T_{c}$=35 K (red), 70 K (yellow), 180 K (red).}
\label{Figure_supp_temp}
\end{figure}
\begin{figure}[h]
\includegraphics[width=0.85\textwidth]{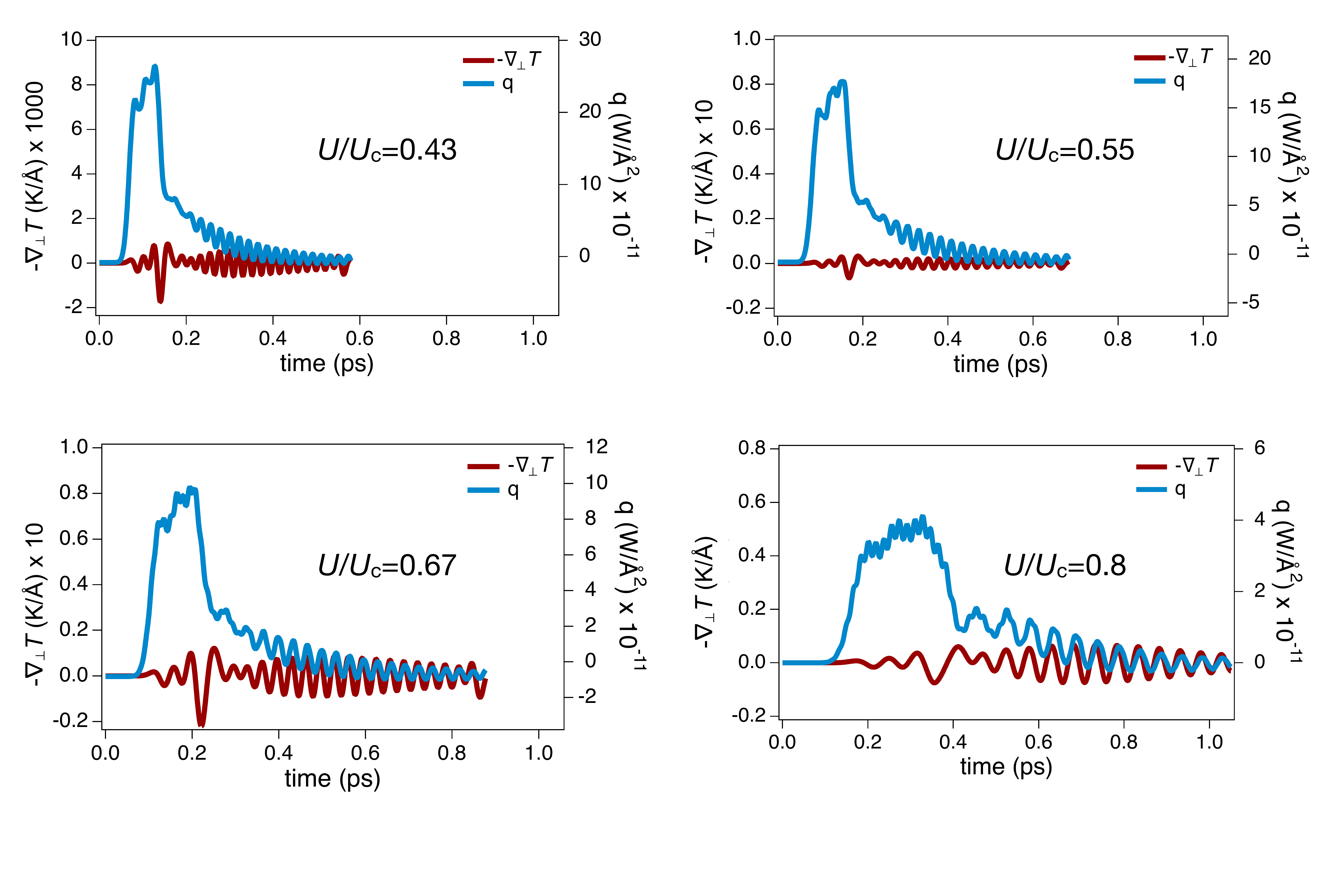}
\caption{Dynamics of the interlayer temperature gradient -$\nabla_{\perp}T(t)$ (purple line) and heat flux $q(t)$ (blue line) at the 15$^{th}$ layer for different values of $U/U_c$. The data reported in the figure have been used to retrieve the electronic thermal conductivity in the Fourier-like regime.}
\label{Figure_supp_conductivity}
\end{figure}
\clearpage

\subsection*{Allowed frequencies and wavevectors}
As discussed in Ref. \citenum{Gandolfi2019}, the wave equation for the temperature variation $\Delta T(t,z)$=$T(t,z)$-$T_{\mathrm{c0}}$ is derived from the combination of Eq. 9  and the local energy conservation. 
A wave-like behaviour of $\Delta T(x,t)$ emerges \cite{Gandolfi2019} when the wavevector $k$ falls in the range $k_{lo}<k<k_{hi}$ with
\begin{equation}
\displaystyle{k_{lo(hi)}=\sqrt{\frac{2}{\alpha \tau_T}\left(\frac{ \tau_q}{ \tau_T}\right)\left(1-\frac{1}{2}\frac{\tau_T}{\tau_q}-(+)\sqrt{1-\frac{\tau_T}{\tau_q}}\ \right)}}
\label{wavevector}
\end{equation}
yielding real and imaginary frequency components:
\begin{flalign}
&\omega_1 =\mp\sqrt{-\left[\frac{\alpha^2}{4}\left(\frac{\tau_T}{\tau_q}\right)^2 k^4+\frac{\alpha}{\tau_q}\left[{1\over2}\left(\frac{\tau_T}{\tau_q}\right) -1\right]k^2+\frac{1}{4\tau_q^2}\right]}\\
&\omega_2=\left[\frac{1}{2\tau_q}+\frac{\alpha}{2}\left(\frac{\tau_T}{\tau_q}\right) k^2\right]
\end{flalign}
The $Q$-factor is easily calculated as $Q$=$|\omega_1|/\omega_2$

\subsection*{Scattering time in SVO from optical data}

The electronic scattering rate can be extracted from optical spectroscopy data. For strongly interacting electrons, the interactions strongly affects the scattering rate and give rise to a frequency dependent scattering rate, which is accounted for by the extended Drude model. The inverse scattering is directly related to the Drude dielectric function through the relation:
\begin{equation}
\label{eq_scattering_time}
\frac{1}{\tau(\omega)}=-\frac{\omega^2_p}{\omega}\mathrm{Im}\frac{1}{\epsilon_D(\omega)-\epsilon_{\infty}}
\end{equation}
$\omega_p$ being the undressed plasma frequency and $\epsilon_{\infty}$ the effective permittivity that accounts for the interband transitions involving electronic states in the valence and other bands.
In Fig. \ref{Figure_tau} we report the optical scattering rate extracted from the optical data reported in Ref. \citenum{Zhang2015}. Considering the density of carriers in SVO, $n$=1.76$\cdot$10$^{22}$ cm$^{?3}$, we obtain $\omega_p$=4.6 eV, which corresponds to the plasma frequency obtained from the fitting with a Drude model.
The calculated $\tau(\omega)$ shows a pronounced decrease in the range 0.1-0.6 eV, which is connected to the strong scattering with modes at very high frequency, in agreeement with the photoemission data discussed in the next session. A value $\tau\approx$2 fs is reached at $\omega >$0.6 eV, whereas the low-frequency limit is $\tau\approx$5 fs. If we consider a pump-probe experiment, which represents a promising configuration to observe wave-like temperature oscillations, it is natural to assume that the high-energy ($>$0.6 eV) excitations, directly photoinjected by the pump pulse (usually in the near-infrared or visible), scatter with high-energy modes within $\approx$1 fs. The low-energy electrons generated during this fast relaxation stage experience a progressively slower scattering time, which reaches a constant value of $\approx$5 fs at energies smaller than 0.1 eV (see Fig. \ref{Figure_tau}). Since thermalization is related to the low-energy carriers, we can assume that the typical thermalization time is $\tau_T \approx$5 fs.
\begin{figure}[t]
\includegraphics[width=0.5\textwidth]{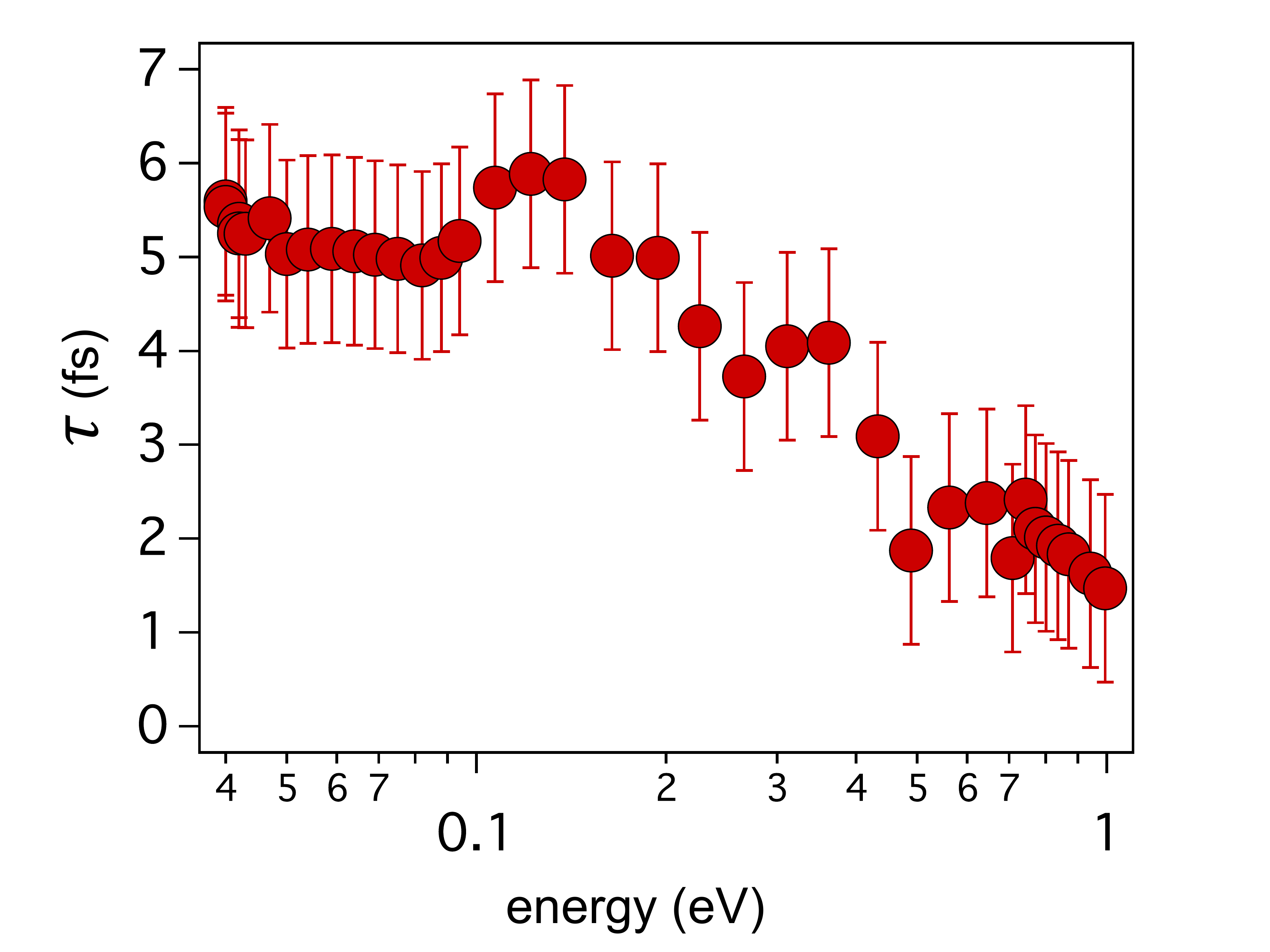}
\caption{Optical scattering time of the conduction electrons for SVO, as extracted from the optical spectroscopy data reported in Ref. \cite{Zhang2015} by using Eq. \ref{eq_scattering_time}.}
\label{Figure_tau}
\end{figure} 
 
\subsection*{Scattering time in SVO from photoemission}
In order to evaluate the local thermalization time we can also refer to the SVO band structure, as measured by angle-resolved photoemission spectroscopy \cite{Aizaki2012}. The electronic correlations manifest themselves in a high-energy contribution to the electronic self-energy, which exhibits a broad kink at $\hbar\Omega_k\sim$0.3 eV. We can thus assume that the typical timescale of electron-electron interactions is 1/$\hbar\Omega_k\sim$2 fs, which is perfect agreement with the high-energy scattering rate extracted from optical data, as discussed in the previous section.

\end{widetext}

\end{document}